\documentclass[%
 aip,
 amsmath,amssymb,
 reprint,%
]{revtex4-1}

\usepackage{graphicx}% Include figure files
\usepackage{dcolumn}% Align table columns on decimal point
\usepackage{bm}% bold math
\usepackage[utf8]{inputenc}
\usepackage[T1]{fontenc}
\usepackage{mathptmx}
\usepackage{etoolbox}
\usepackage{comment}
\usepackage{subcaption}
\makeatletter
\def\@email#1#2{%
 \endgroup
 \patchcmd{\titleblock@produce}
  {\frontmatter@RRAPformat}
  {\frontmatter@RRAPformat{\produce@RRAP{*#1\href{mailto:#2}{#2}}}\frontmatter@RRAPformat}
  {}{}
}%
\makeatother
\begin{document}

\preprint{AIP/123-QED}

\title[]{Effect of Porous Coating on the Water-Exit Dynamics of a Circular Cylinder}
% Force line breaks with \\
\author{Intesaaf Ashraf}
\email{intesaaf.ashraf@ucl.ac.uk}

\affiliation{
PtYX Lab, UR-CESAM, D\'epartement de Physique B5,
Universit\'e de Li\`ege, Li\`ege 4000, Belgium
}

\altaffiliation{
Present address: Department of Mechanical Engineering,
University College London, London, United Kingdom
}

\author{Mahender Thotakuri}
\affiliation{
Department of Mechanical \& Aerospace Engineering,
Indian Institute of Technology Hyderabad,
Sangareddy 502285, Telangana, India
}

\author{Neetu Tiwari}
\affiliation{
Department of Mechanical \& Aerospace Engineering,
Indian Institute of Technology Hyderabad,
Sangareddy 502285, Telangana, India
}

\author{Stephane Dorbolo}
\affiliation{
PtYX Lab, UR-CESAM, D\'epartement de Physique B5,
Universit\'e de Li\`ege, Li\`ege 4000, Belgium
}
\date{\today}% It is always \today, today,
             %  but any date may be explicitly specified

\begin{abstract}

The effect of a porous coating on the water-exit dynamics of a circular cylinder is investigated experimentally. Cylinders comprising a $30~\mathrm{mm}$ solid core surrounded by a $10~\mathrm{mm}$-thick porous foam layer with pore densities of 10, 20, and 60 pores per inch (PPI) are compared with a smooth cylinder of the same outer diameter. Experiments are performed over $1.50\times10^{4}\leq Re\leq5.00\times10^{4}$ and $0.37\leq Fr\leq4.08$. Time-resolved particle image velocimetry characterizes the wake and reconstructed pressure field, while proper orthogonal decomposition (POD) examines its energetic organization. All porous cylinders reduce free-surface elevation relative to the smooth cylinder, with the largest reduction generally obtained for 60 PPI. The submerged wake, however, is not simply weakened. Porous cases generally develop larger circulation and vortex area, with a non-monotonic dependence on pore density and operating condition. The 10 PPI cylinder frequently produces the largest vortex region, whereas 60 PPI generates stronger negative pressure within the vortex and the largest integrated vortex-associated pressure deficit for $Fr=1.02$, $2.00$, and $4.08$. During the submerged approach to contact, this pressure deficit co-evolves strongly with circulation, with Pearson correlation coefficients typically between $0.90$ and $0.98$. POD shows that 9--18 modes recover 95\% of the modal energy, with 20 PPI requiring only 9--10 modes. Overall, porous coating reorganizes wake strength, extent, pressure signature, and energetic structure while reducing the free-surface response.

\end{abstract}

\maketitle

\section{\label{intro}Introduction}

The entry and exit of a  body through a liquid--air interface are 
classical free-surface problems with applications in naval architecture, 
underwater vehicles, ocean engineering, aerospace systems and biological 
locomotion. Compared with water entry, however, the physics of water exit 
has received considerably less attention. Water-entry studies have examined, 
among other aspects, the influence of body geometry, impact velocity and 
structural response in considerable detail 
\cite{challa2014,challa2010,mohtat2015,yang2012}. Water exit is in many 
respects a different problem. As a submerged body approaches and crosses the 
free surface, the surrounding liquid must be displaced, the interface is 
strongly deformed, a part of the liquid can be carried out of the bath, and 
the wake generated beneath the body interacts with the deforming interface. 
The relative importance of inertia, gravity, viscosity, surface tension and 
body geometry can therefore change considerably during a single exit event 
\cite{Havelock1936,truscott2016,wu2017experimental,haohao2019numerical}.

One of the early contributions to this problem was made by Havelock 
\cite{Havelock1936}, who developed an analytical description of the forces 
associated with a circular cylinder moving in a fluid. Greenhow and Lin 
\cite{greenhow1983nonlinear} later investigated the rise of a neutrally 
buoyant cylinder experimentally and theoretically. They showed that the 
upward motion of the body deforms the initially flat free surface into a 
pronounced mound or bump. At sufficiently large deformation this structure 
breaks down in an irregular manner, which they referred to as ``waterfall 
breaking''. Telste \cite{telste1987} subsequently studied the approach of a circular cylinder 
toward a free surface using potential-flow theory and showed that the nature 
of the interaction changes considerably with the speed of the body. These 
early studies established that the water-exit problem cannot be understood 
only from the motion of the  body; the deformation and eventual breakup 
of the free surface form an essential part of the dynamics.

The surface surge associated with water exit was investigated further by 
Liju et al. \cite{liju2001}, who considered axisymmetric bodies moving 
normal to the interface and compared experiments with boundary-element 
calculations. Their study provided useful information on the formation and 
evolution of the free-surface surge for different body geometries and 
Reynolds numbers. Numerical approaches were subsequently developed to deal 
with the large interface deformation involved in this problem. Greenhow and 
Moyo \cite{greenhow1997water} studied the entry and exit of horizontal circular cylinders, while 
Kleefsman et al. \cite{kleefsman2004improved} and Nair and Bhattacharyya 
\cite{nair2018water} used Volume-of-Fluid-based approaches for problems 
involving strongly deforming free surfaces. \citet{moshari2014numerical} extended this type 
of analysis to two- and three-dimensional water exit of circular cylinders, 
showing wave generation and air entrainment during the exit process. Chu 
et al. \cite{chu2010} considered a cylinder moving in its longitudinal 
direction and observed cavities at the ends of the body. Collapse of these 
cavities resulted in water slapping and large local pressure variations, 
with different pressure responses for accelerating and decelerating motion.

The influence of body geometry was also investigated using spheroidal and 
spherical bodies. \citet{ni2015} simulated the complete exit of a 
fully submerged spheroid and showed that the detachment of the free surface 
from the body can be delayed by increasing the bluntness of the body.  \citet{haohao2019numerical} used the lattice Boltzmann method to 
investigate a sphere exiting at constant velocity. Their results showed a 
strong dependence of the maximum free-surface elevation on the Froude number 
at relatively low Froude numbers, whereas this dependence became weaker at 
larger values. They also observed more violent waterfall breaking as the 
Froude number increased and found that the Reynolds number plays an important 
role while the sphere remains fully submerged.

Experiments on freely rising spheres provided further information about the 
connection between the submerged wake and the subsequent water-exit process. 
\citet{truscott2016} investigated buoyant spheres rising 
toward and through the free surface. Depending on the release depth and 
Reynolds number, the sphere could follow either an approximately straight 
trajectory or an oscillatory path, with the change in trajectory being 
associated with the development and shedding of wake vortices. 
\citet{wu2017experimental} investigated both fully and partially submerged 
spheres and showed that the free-surface elevation increases with the sphere 
velocity, or equivalently with the Froude number. Once the sphere becomes 
partially emerged, the liquid carried with it evolves into a water column, 
whose extent is also strongly dependent on the exit velocity.

More recent experiments have started to examine the complete sequence from 
the submerged motion to interface crossing and drainage rather than focusing 
only on the maximum surface elevation.  
\citet{ashraf2024experimental} investigated horizontal circular cylinders exiting  water and silicone oil at constant velocity. They showed that the initial submergence depth no longer has a significant influence on the interface deformation once the cylinder starts sufficiently far below the surface. The wake behaved differently in the two liquids: symmetry breaking developed in water, whereas the wake remained comparatively symmetric in the more viscous silicone oil. The thickness of liquid above the cylinder at the 
beginning of interface crossing increased approximately logarithmically with 
the Froude number. After crossing, the entrained liquid first drained 
approximately exponentially before entering a later regime in which the film 
thickness varied as the inverse square root of time. Interestingly, this 
late drainage regime was largely independent of the crossing velocity. 
These measurements also showed that a large cylinder aspect ratio together 
with end plates is required when the objective is to approach a 
two-dimensional flow configuration.

The effect of a non-circular geometry was investigated by Ashraf and Dorbolo 
\cite{ashraf2024square} using a square cylinder pulled vertically through 
the free surface at constant speed. As for circular and spherical bodies, 
the free surface began to rise before the body reached it, and the maximum 
surface elevation increased with Froude number. Synchronized force 
measurements allowed the submerged drag, interface cross-over force and 
entrainment force to be separated. The drag and entrainment forces increased 
with increasing velocity, whereas the net cross-over force decreased. PIV 
measurements showed a pair of clockwise and counter-clockwise vortices 
attached to the cylinder, with separation beginning at the sharp leading 
edges. No classical von K\'arm\'an vortex street was observed over the 
range of conditions investigated.

The role of this wake was examined in greater detail in a subsequent study 
of the same square-cylinder configuration \cite{ashraf2025wake}. 
Time-resolved PIV showed that the wake is dominated by a persistent pair of 
counter-rotating vortices rather than periodic vortex shedding. The 
circulation exhibited two regimes: it increased rapidly with Froude number 
below approximately $Fr \simeq 1$ and approached a much weaker dependence 
above this value. This change was consistent with the two-regime behaviour 
previously observed for the entrainment force. At the same time, the area 
occupied by the vortices changed only weakly with Froude number, whereas 
their swirl strength and the extent of the rotation-dominated region 
increased. This suggests that the increase in liquid entrainment is related 
more closely to the strengthening of the vortical motion than simply to an 
increase in vortex size.

Most of the studies discussed above considered a nearly constant crossing 
velocity. The role of acceleration has received much less attention. 
Ashraf and Dorbolo \cite{ashraf2026experimental} recently investigated a 
circular cylinder accelerating vertically from rest and used time-resolved 
PIV to follow the starting vortex generated beneath the cylinder. The vortex 
trajectory, equivalent radius, peak vorticity and circulation were measured 
for different imposed accelerations. A collapse of the circulation data was 
obtained using an impulse-based scaling involving acceleration and cylinder 
displacement. The scaling remained valid over much of the submerged motion, 
but a consistent change occurred as the vortex approached the free surface, 
showing directly that the interface modifies the vortex evolution. The 
vortex trajectories nevertheless remained predominantly vertical over the 
range of accelerations considered.

Surface condition provides another means of modifying water-exit dynamics. 
Ashraf and Dorbolo \cite{ashraf2024effect} compared smooth and dimpled 
spheres exiting water at a constant velocity. At low Froude numbers, the 
behaviour of the two spheres was similar, but at larger Froude numbers, the 
dimpled sphere produced a smaller maximum surface elevation and entrained 
less liquid. The dimpled sphere consequently showed lower drag and 
entrainment coefficients, while the cross-over force coefficient remained 
approximately unchanged. These observations demonstrate that relatively 
small modifications of the body surface can alter the wake and, through it, 
the amount of liquid carried through the interface.

A related question is whether fluid passage through the body itself can 
modify the exit dynamics. \citet{takamure2025motion} 
investigated a sphere containing a single axial through-hole and varied its 
initial submergence depth. Increasing the submergence depth increased the 
amount of entrained water and the kinetic-energy loss of the sphere. At the 
largest depth, the sphere rotated while crossing the interface and produced 
a sheet-like water mass aligned with the through-hole. The vented sphere 
also lost more energy than a  sphere at a comparable Reynolds number. 
This result is particularly important because it shows that allowing fluid 
to pass through an object can influence not only the local flow but also 
entrainment, rotation and the subsequent trajectory of the body.

At higher exit velocities, cavity dynamics become increasingly important. 
\citet{zheng2025experimental} experimentally studied slender 
cylinders over a range of exit velocities, head shapes, length-to-diameter 
ratios and elastic moduli. Increasing the velocity enlarged the cavity and 
reduced the measured drag coefficient, although the cavity became less 
stable at the highest velocities. Conical heads produced lower drag and 
more regular cavity shapes than the truncated-cone configurations, while an 
increase in the length-to-diameter ratio also reduced drag. Their experiments 
further showed that structural stiffness can become important: relatively 
flexible bodies underwent substantial bending and trajectory changes, 
whereas the stiffer bodies remained close to their original shape during 
exit. Thus, at sufficiently high speed, geometry, cavity dynamics and 
structural deformation cannot be considered independently.

The water-exit problem becomes still more complicated when the initially 
flat free surface is replaced by a wave. \citet{zhou2024numerical} numerically investigated a high-speed projectile 
crossing a wave and showed that the cavity develops asymmetrically because 
its expansion toward the free surface differs from that into the surrounding 
water. This asymmetry generates differences in pressure between the upper 
and lower surfaces of the projectile and consequently produces lift and 
pitching moments. Changing the vertical location of the trajectory relative 
to the wave changed the cavity opening, wave deformation and resulting 
loads. \citet{huang2024numerical} considered the lower-speed 
retrieval of a deep-sea mining vehicle and similarly found that the wave 
phase strongly affects the attitude and horizontal drift of the vehicle. 
The lifting force must therefore be chosen together with the wave phase; 
simply increasing the lifting force or retrieval velocity does not 
necessarily improve the recovery process.

More recently, \citet{zhou2026aerial} examined the complete 
water-entry--underwater--water-exit sequence of a cylindrical body crossing 
a wave and identified two different modes depending on whether cavity 
closure occurs before or after the body begins to exit. In the sequential 
mode, the cavity closes before water exit and its collapse generates an 
internal jet and periodic load fluctuations. In the simultaneous mode, the 
body begins to exit while the cavity is still open, leaving a one-sided 
cavity connected to the atmosphere. The resulting cavity topology, pressure 
distribution and hydrodynamic loads were therefore strongly dependent on 
the phase at which the body entered the wave. These results show that in a 
wave environment the entry and exit stages cannot always be treated as two 
independent events.

Environmental objects near the free surface introduce another level of 
complexity. \citet{wang2025cfd} used a coupled CFD--FEM approach 
to investigate a ventilated vehicle exiting through floating ice. Contact 
with the ice distorted the cavity, promoted its collapse, increased wetting 
of the vehicle and produced stronger splashing and vortex-dominated regions. 
Ice impact also increased kinetic-energy dissipation and reduced trajectory 
stability. The response depended strongly on the impact location, with 
off-centre impacts producing repeated collisions, deflection of the vehicle 
and an asymmetric tail cavity. Complementary experiments by \citet{zhang2026experimental} considered impacts with floating obstacles during  water exit. They also observed asymmetric cavity collapse, strong pressure pulsations and local structural deformation. The initial relative position of the obstacle was found to be particularly important and, in some cases, more important than its detailed shape.

Alongside these experiments and continuum CFD calculations, numerical 
methods for moving irregular objects across gas--liquid interfaces are also 
developing. \citet{yong2025coupled} introduced a coupled 
particle-dynamics/lattice-Boltzmann framework in which irregular bodies are 
represented using a multi-sphere description. The method reproduced 
benchmark problems including particles entering and being pulled out of a 
liquid and provides a useful route for resolving moving-contact-line and 
gas--liquid-body interactions for non-standard geometries, although the 
present formulation remains limited to two-dimensional or pseudo-three-
dimensional configurations.

The literature therefore shows that water exit is governed by a strong
coupling between body motion, wake development, free-surface deformation,
liquid entrainment and drainage. The Froude and Reynolds numbers remain
important parameters, but geometry, acceleration, surface condition,
submergence depth, structural flexibility and the state of the free surface
can all modify the process. Nevertheless, one aspect remains largely
unexplored: the water exit of bodies covered by a distributed porous layer.
Existing studies have shown that surface dimples can modify separation and
entrainment and that a single through-hole can alter the carried water mass,
energy loss and trajectory. Neither configuration, however, represents a
body whose outer surface contains a finite-thickness porous layer into which
the surrounding liquid can penetrate. Such a porous coating can modify the
near-surface pressure and velocity fields, alter the development of the
separated shear layers, and consequently reorganize the wake and its
interaction with the free surface. The present work addresses this gap by
investigating how a porous coating modifies the water-exit dynamics of a
circular cylinder, with particular attention to the wake and free-surface
response.

\section{METHODOLOGY}
\subsection{\label{sec:setup}Experimental setup}

The experimental facility consisted of a vertical lifting mechanism, a glass water tank, a time-resolved particle image velocimetry (PIV) system, and the test cylinder. The internal dimensions of the water tank were $78.5~\mathrm{cm}\times27.5~\mathrm{cm}\times72.5~\mathrm{cm}$. The lifting mechanism employed a rack-and-pinion system connected to a rigid carbon-fibre support frame, which held the cylinder horizontally and maintained its orientation during the upward motion. Further details of the lifting system are provided in our previous work \citep{ashraf2024experimental, ashraf2025wake, ashraf2024square}.

The test cylinder had an outer diameter of $D=50~\mathrm{mm}$ and a spanwise length of $L=360~\mathrm{mm}$, corresponding to an aspect ratio $L/D=7.2$. The porous cylinders consisted of a solid cylindrical core of diameter $30~\mathrm{mm}$ surrounded by a $10~\mathrm{mm}$-thick annular layer of porous foam, giving an overall cylinder diameter of $D=50~\mathrm{mm}$. The foam was attached uniformly around the circumference of the solid core. Three foam grades were investigated, with pore densities of 10 (coarse pores), 20 (medium pores), and 60 (very fine pores) pores per inch (PPI). Here, PPI denotes the nominal number of pores per inch of the foam and is used solely to distinguish the different porous-surface configurations. The porosity and permeability of the foam layers were not independently measured. Accordingly, comparisons among the 10, 20, and 60 PPI cases are interpreted as differences between porous-surface configurations rather than as a direct parametric dependence on porosity or permeability.

The outer diameter and spanwise length were kept identical for all porous cases and for the smooth reference cylinder, so that the projected geometry remained unchanged between experiments. Although the outer dimensions and projected geometry were identical, the porous annulus is accessible to the surrounding liquid and therefore does not necessarily displace liquid in the same manner as the smooth reference cylinder. The effective hydrodynamically displaced volume may therefore differ between the smooth and porous configurations. The present measurements do not quantify the amount of liquid penetrating or residing within the foam during the motion.

End plates were used at both ends of the cylinder to reduce
finite-span effects and promote an approximately two-dimensional flow
in the central region. The PIV measurement plane passed through the
cylinder mid-span, where the influence of the end boundaries was
expected to be smallest. The cylinder was initially submerged with its centre approximately $326~\mathrm{mm}$ below the undisturbed free surface. During the upward motion, the instantaneous dimensionless immersion depth was defined as $h=H_i/D$, where $H_i$ is the vertical distance between the cylinder centre and the undisturbed free surface. With this definition, $h=0.5$ defines the nominal contact position, at which the upper surface of the cylinder reaches the level of the initially undisturbed free surface. Because the interface deforms during the approach, this reference position does not necessarily coincide with the first physical contact between the cylinder and the instantaneous free surface.

\begin{figure*}[t]
    \centering
    \includegraphics[width=\textwidth]{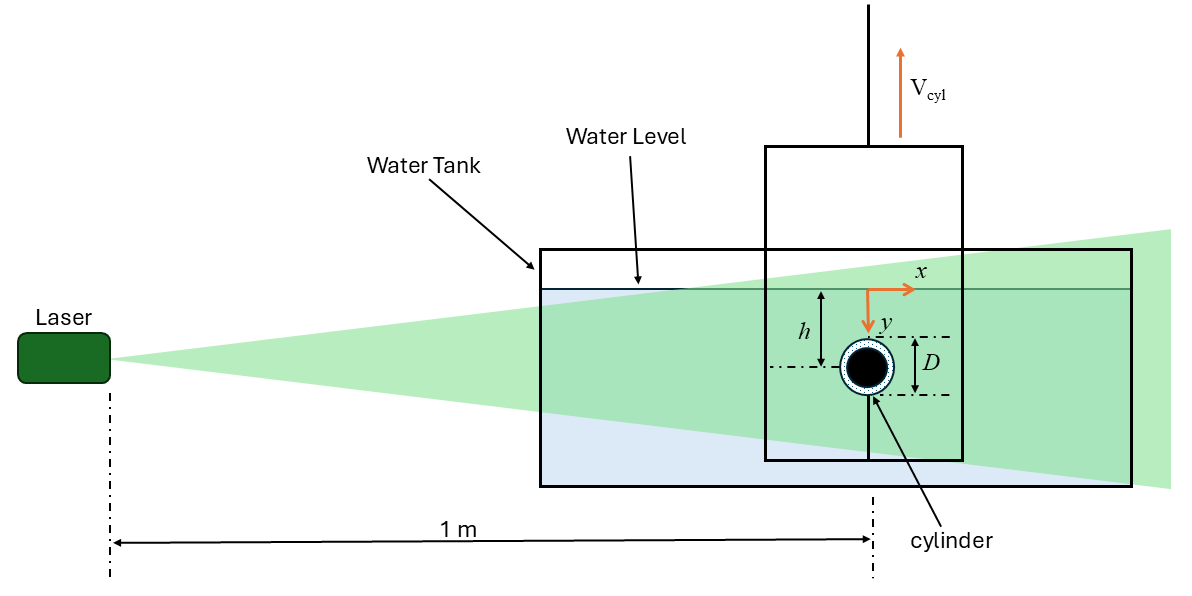}
    \caption{Schematic of the experimental arrangement used to investigate the water-exit dynamics of a porous horizontal circular cylinder. The cylinder is translated vertically upward with velocity $V_{\mathrm{cyl}}$ through the initially undisturbed free surface. A laser sheet illuminates the $x$--$y$ measurement plane passing through the cylinder mid-span. The laser source is positioned approximately $1~\mathrm{m}$ from the cylinder, $D$ denotes the cylinder diameter, and $h$ represents the instantaneous dimensionless immersion depth.}
    \label{fig:experimental_setup}
\end{figure*}

The coordinate system is shown in Fig.~\ref{fig:experimental_setup}. The $x$-axis is horizontal, while the $y$-axis is directed vertically downward. The origin is located on the cylinder pulling axis at the initially undisturbed free surface, so that the instantaneous cylinder position used in the analysis is referenced directly to the free-surface level.

Two-dimensional time-resolved PIV was used to characterize the flow in the $x$--$y$ plane. The flow was seeded with tracer particles of approximately $20~\mu\mathrm{m}$ diameter and illuminated using a $532~\mathrm{nm}$ laser. Images were acquired at $2000~\mathrm{Hz}$ and processed using \textit{PIVlab} \citep{thielicke2014pivlab}, with a final interrogation-window size of $32\times32$ pixels. The measurement plane passed through the central span of the cylinder. Standard vector-validation and smoothing procedures available in \textit{PIVlab} were applied to remove spurious vectors and reduce local measurement noise. The uncertainty in the velocity measurements was estimated using the approach of \citet{sciacchitano2013piv}, giving typical absolute uncertainties of $0.03$--$0.08~\mathrm{m\,s^{-1}}$, corresponding to approximately $3$--$5\%$ of the local instantaneous
velocity in the dynamically active regions.

Experiments were performed at prescribed terminal cylinder velocities
of $0.30$, $0.50$, $0.70$, and $1.00~\mathrm{m\,s^{-1}}$.
The cylinder was accelerated from rest at $4~\mathrm{m\,s^{-2}}$
until the required terminal velocity was reached, and the motion was
verified to remain approximately constant over the PIV measurement
interval. One time-resolved PIV realization was acquired for each
surface--velocity condition. Consequently, the instantaneous PIV
samples within each record represent the temporal evolution of a
single experimental realization rather than independent experimental
repeats.

The Reynolds and Froude numbers are defined as

\begin{equation}
Re=\frac{UD}{\nu},
\end{equation}

and

\begin{equation}
Fr=\frac{U^2}{gR},
\end{equation}

where $U$ is the terminal cylinder velocity, $D=50~\mathrm{mm}$ is
the cylinder diameter, $R=D/2$ is its radius, $\nu$ is the kinematic
viscosity of water, and $g$ is the gravitational acceleration.
The four test velocities correspond to
$Re=1.50\times10^4$, $2.50\times10^4$, $3.50\times10^4$, and
$5.00\times10^4$, and to $Fr=0.37$, $1.02$, $2.00$, and $4.08$,
respectively. Since both $Re$ and $Fr$ vary with the prescribed terminal velocity $U$,
the two nondimensional parameters were not varied independently in the
present experiments. Increasing the exit velocity therefore changes
$Re$ and $Fr$ simultaneously. Consequently, trends across the four
operating conditions should not be interpreted as isolated
Froude-number or Reynolds-number effects, and the present dataset does
not permit their individual contributions to be separated.

\subsection{Vortex and pressure quantities}

The pressure field was reconstructed from the time-resolved PIV
measurements using the pressure-Poisson procedure described in our
previous work~\cite{ashraf2025wake}. The numerical formulation,
boundary conditions, uncertainty considerations, and definition of the
pressure coefficient $C_p$ follow that procedure. Since the PIV
measurements do not resolve the velocity within the porous material,
the velocity inside the cylinder region, including the porous annulus,
was set to zero prior to the pressure reconstruction, following the
treatment used for the smooth cylinder in Ref.~\citenum{ashraf2025wake}.
The same masking procedure was applied consistently to all four surface
configurations. Consequently, the reconstructed pressure field
describes the external flow and does not resolve the pressure or flow
within the porous layer. The pressure quantities reported here are
evaluated over the identified external wake-vortex region,
$\Omega_v$, which lies outside the masked cylinder region.

The vortex was identified using the second-moment procedure described
and validated in our previous work \cite{ashraf2026experimental},
adapted from the method of Steiner et al. \cite{steiner2023vortex}.
Briefly, the spanwise-vorticity field was thresholded using a
threshold coefficient of 0.15, and the largest connected region of the
selected vorticity sign was retained as the vortex region. The
sensitivity of the identified vortex properties to the threshold value
was examined previously over the range 0.10--0.25 and was found to be
small \cite{ashraf2026experimental}.

The wake consists of an approximately symmetric pair of
counter-rotating vortices beneath the cylinder. Since the two vortices
showed comparable evolution, the clockwise vortex was selected as a
representative structure and was used consistently for all quantitative
comparisons reported here. The black contour shown in
Fig.~\ref{fig:wake_pressure_evolution_Re35000} delineates the
identified clockwise vortex region, $\Omega_v$, whose area is
denoted by $A_v$. In the present analysis, $A_v$ denotes the actual
area of the retained connected vortex mask rather than the equivalent
area $\pi r_{\mathrm{eq}}^2$ obtained from the second moments.

The vortex circulation and area are reported in nondimensional form
as $|\Gamma|/(UD)$ and $A_v/D^2$, respectively, where $\Gamma$ and
$A_v$ denote the circulation and area of the identified clockwise
vortex. The vorticity fields are presented in nondimensional form as
$\omega D/U$.

To characterize the pressure associated with the vortex, the mean
pressure coefficient within the identified vortex region is defined as

\begin{equation}
\overline{C}_{p}^{\,v}
=
\frac{1}{A_v}
\int_{\Omega_v} C_p\,\mathrm{d}A .
\end{equation}

A positive-definite measure of the mean pressure deficit is defined as

\begin{equation}
\Pi_v
=
\frac{1}{A_v}
\int_{\Omega_v}
\max(-C_p,0)\,\mathrm{d}A .
\end{equation}

Thus, only the negative-pressure contribution within the vortex region
is retained in $\Pi_v$, thereby avoiding cancellation by local positive
values of $C_p$.

The corresponding integrated vortex-associated pressure deficit is
defined as

\begin{equation}
I_{p,v}
=
\frac{1}{D^2}
\int_{\Omega_v}
\max(-C_p,0)\,\mathrm{d}A ,
\end{equation}

which may equivalently be written as

\begin{equation}
I_{p,v}
=
\Pi_v\frac{A_v}{D^2}.
\end{equation}

Unlike $\Pi_v$, which represents the mean intensity of the negative
pressure within the vortex region, $I_{p,v}$ also accounts for the
spatial extent of that region and is therefore used as an integrated
measure of the vortex-associated pressure deficit.

To obtain a representative wake state around the nominal contact
position without relying on a single instantaneous PIV frame, the
wake quantities were averaged over the narrow symmetric interval
$0.45\leq h\leq0.55$, centred on $h=0.5$. Since $D=50~\mathrm{mm}$,
this interval corresponds to a cylinder displacement of $\pm0.05D$
($\pm2.5~\mathrm{mm}$) about the nominal contact position. The
corresponding error bars denote the standard deviation of the
instantaneous values within this positional interval and therefore
characterize temporal variability within a single PIV realization
rather than run-to-run experimental repeatability.

\subsection{Proper orthogonal decomposition}

To complement the vortex-based characterization, proper orthogonal
decomposition (POD) was applied to the time-resolved PIV velocity
fields for each surface--velocity condition. The POD modes were
ordered according to their associated eigenvalues, $\lambda_i$, which
measure the contribution of each mode to the total fluctuation energy.
The cumulative energy retained by the first $r$ modes was defined as

\begin{equation}
E_r=
\frac{\displaystyle\sum_{i=1}^{r}\lambda_i}
{\displaystyle\sum_{i=1}^{N}\lambda_i},
\end{equation}

where $N$ is the total number of available POD modes. The quantities
$r_{90}$, $r_{95}$, and $r_{99}$ denote the minimum numbers of modes
required to recover 90\%, 95\%, and 99\% of the total POD energy,
respectively.

Because a separate decomposition was performed for each
surface--velocity condition, the modal-energy distributions are used
here primarily to compare the dimensionality of the measured wake
response. Individual modes from different decompositions are not
assumed to have a one-to-one correspondence solely on the basis of
their mode number. As only one time-resolved PIV realization was
available for each condition, the POD quantities are interpreted as
descriptive measures of the organization of each measured wake rather
than as estimates of run-to-run statistical variability.

\section{\label{sec:results}Results and Discussion}

\subsection{Free-surface response}

\begin{figure}[t]
    \centering

    \begin{subfigure}[b]{1\linewidth}
        \centering
        \includegraphics[width=\linewidth]{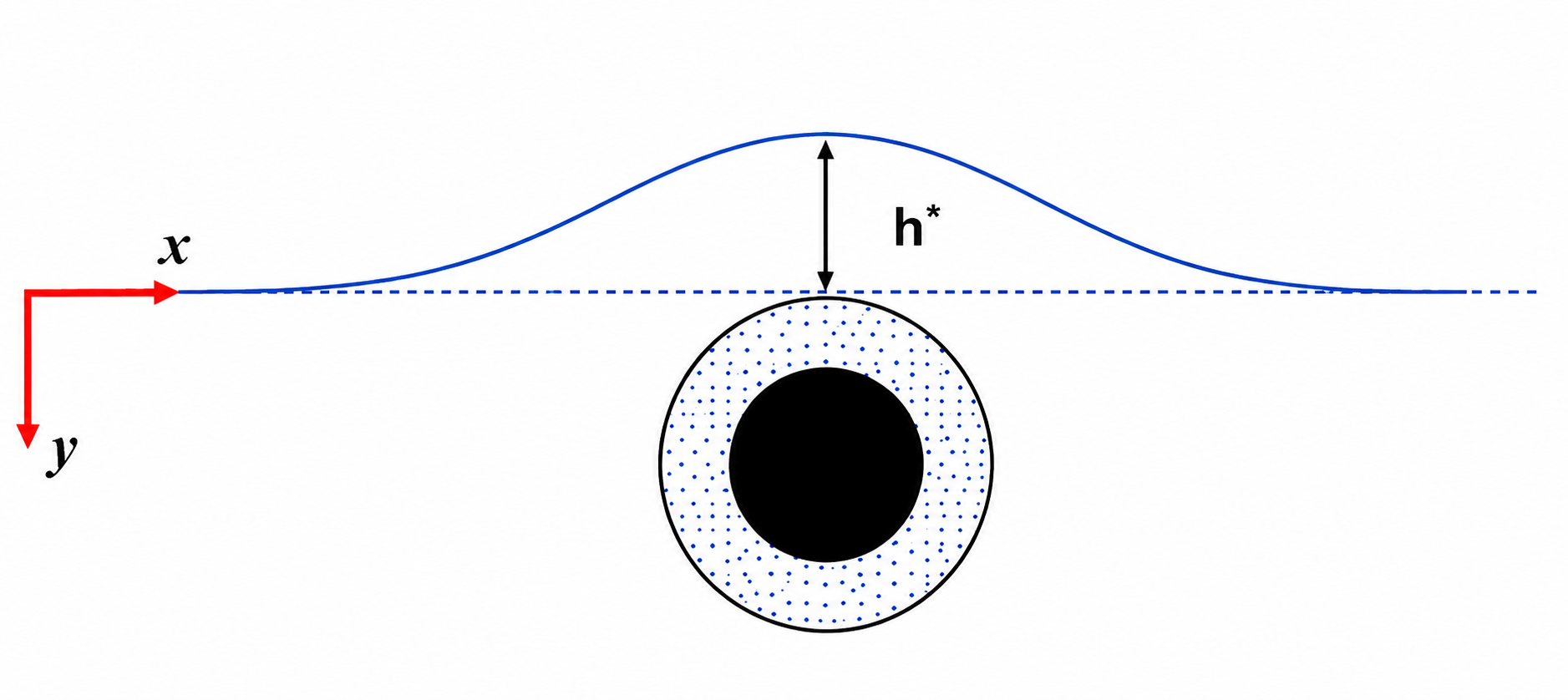}
        \caption{}
        \label{fig:hstar_definition}
    \end{subfigure}
    \hfill
    \\
    \begin{subfigure}[b]{1\linewidth}
        \centering
        \includegraphics[width=\linewidth]{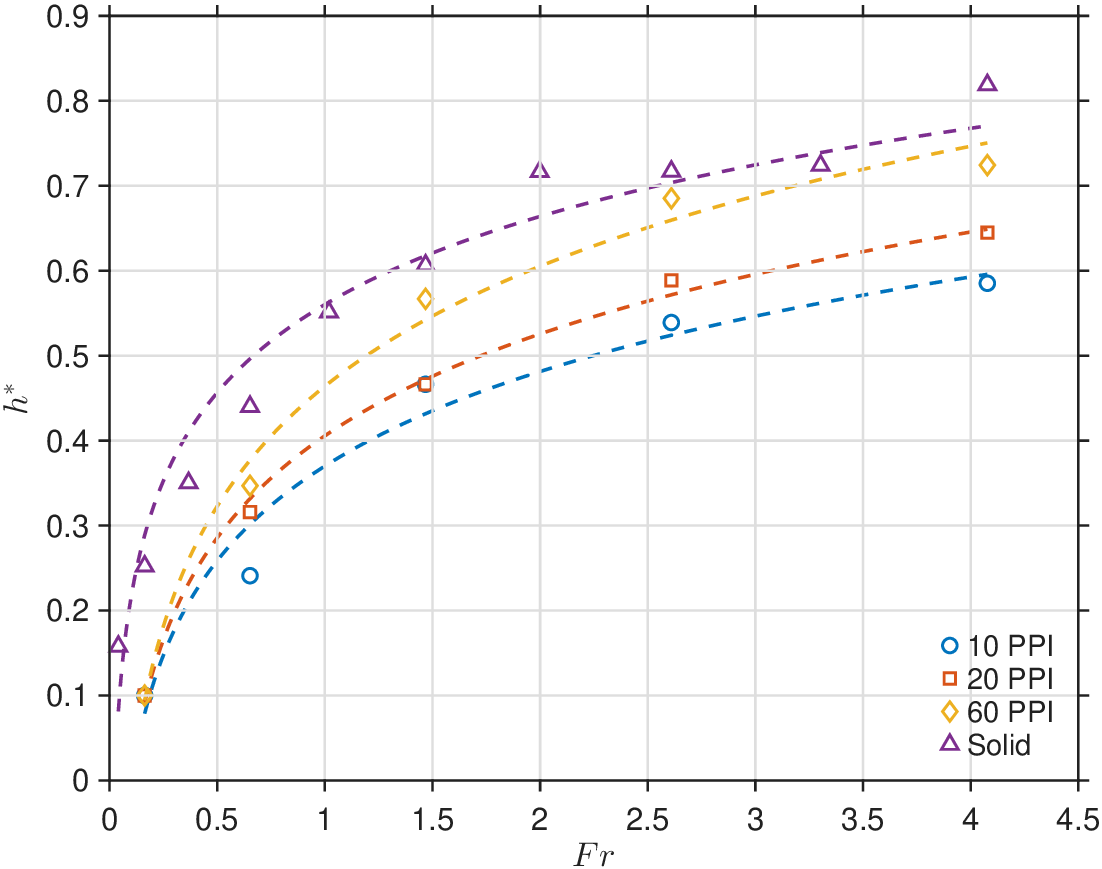}
        \caption{}
        \label{fig:hstar_results}
    \end{subfigure}

    \caption{(a) Schematic illustration of the free-surface elevation used to define the characteristic nondimensional height $h^{*}=\eta/D$ at the nominal contact position, $h=0.5$, where $\eta$ is the free-surface elevation above the initially undisturbed free-surface level. (b) Variation of $h^{*}$ with Froude number $Fr$ for the smooth cylinder and porous cylinders with pore densities of 10, 20, and 60 PPI. Symbols represent the measurements and dashed lines indicate logarithmic fits.}
    \label{fig:hstar_combined}
\end{figure}

The effect of the porous surface is already apparent before the cylinder crosses the interface. Figure~\ref{fig:hstar_combined}(a) defines the free-surface elevation $h^{*}$ measured when the rising cylinder reaches the initially undisturbed free-surface level. As shown in Fig.~\ref{fig:hstar_combined}(b), $h^{*}$ increases rapidly with $Fr$ at the lower velocities and then increases more gradually as $Fr$ becomes larger. This behaviour is reasonably represented by the logarithmic fits shown in the figure and is consistent with the general increase of free-surface deformation with exit velocity reported previously for smooth bodies \cite{wu2017experimental, haohao2019numerical, ashraf2024experimental}.

A clear difference, however, is observed between the smooth and porous cylinders. The smooth cylinder produces the largest surface elevation throughout the range considered, whereas all three porous cases give a lower $h^{*}$. The reduction is most pronounced for the 60 PPI cylinder, with the 10 and 20 PPI cases lying between the smooth and 60 PPI results.  The separation between the different surfaces persists as $Fr$ increases. Thus, the effect of the porous surface is not restricted to the short interval during which the body cuts through the interface. The reduced free-surface elevation of the porous cylinders is accompanied by substantial changes in the submerged wake, as examined below.

This observation motivates examination of the wake immediately below the cylinder. In particular, it is useful to determine whether the lower surface elevation of the porous cylinders is accompanied simply by a weaker wake or whether the structure of the wake itself is reorganized.

\subsection{Wake development during the approach to the free surface}

Representative instantaneous fields at $Re=3.50\times10^{4}$ and $Fr=2.00$ are shown in Fig.~\ref{fig:wake_pressure_evolution_Re35000}. For all four cylinders, a pair of oppositely signed vortical regions develops beneath the body. The associated pressure field contains corresponding regions of negative pressure on either side of the wake. The main difference between the cases is not the existence of these structures but their size, strength, and persistence as the cylinder approaches the interface.

At $h\simeq1.5$, the wake of the smooth cylinder is relatively compact, whereas the porous-cylinder wakes extend farther below the body. This distinction becomes clearer as $h$ decreases. By $h\simeq0.5$, immediately before the nominal contact position, the identified vortex region remains appreciably larger for the porous cylinders than for the smooth cylinder. At the same time, the pressure fields show that a larger vortex does not necessarily correspond to the strongest local pressure deficit. The 60 PPI case, for example, develops a particularly intense negative-pressure region although its identified vortex area is smaller than that of the 10 PPI case. This difference between vortex size and pressure intensity becomes important in the quantitative results below.

The evolution of circulation is shown in Fig.~\ref{fig:gamma_vs_h}. Since $h$ decreases as the cylinder rises, the curves are read from left to right in the direction of motion. In all cases, $|\Gamma|/(UD)$ initially grows as the cylinder moves upward, showing the progressive accumulation of circulation in the wake. The effect of the surface condition is already large well before the cylinder reaches the interface.

At the lowest tested condition, $Fr=0.37$, the 10 PPI cylinder produces the largest circulation over most of the submerged trajectory. At the nominal contact position, $h=0.5$, its value is approximately $2.53$, compared with $2.28$ for 20 PPI, $2.00$ for 60 PPI, and only $1.32$ for the smooth cylinder. The ordering changes as the velocity increases. At $Fr=1.02$, the 60 PPI case reaches a circulation comparable to, and locally larger than, that of the 10 PPI case close to the free surface. At $Fr=2.00$, the 10 and 20 PPI cases are again similar at the nominal contact position, while at $Fr=4.08$ the 20 PPI cylinder has the largest circulation at the nominal contact position, approximately $2.16$. The corresponding values for the 60 PPI, 10 PPI, and smooth cylinders are approximately $1.93$, $1.71$, and $1.36$, respectively.

The response to pore density is therefore not monotonic. Increasing the PPI does not continuously increase or decrease the circulation. Instead, the surface that produces the strongest circulation changes with $Fr$. This is different from a simple picture in which the porous coating merely
weakens the separated wake. The porous surface changes the development of the wake, but the result depends on both the pore configuration and the operating condition.

The change after the nominal contact position, $h=0.5$, is also worth noting. Once $h<0.5$, the circulation curves become less regular and the differences between cases can increase rapidly. This is particularly evident for the 60 PPI cylinder at the higher Froude numbers. At this stage the wake is no longer evolving beneath an effectively unbounded liquid region: it is interacting directly with the deforming interface while the wetted portion of the cylinder is simultaneously changing. The departure of the circulation from its earlier trend is therefore consistent with the free surface beginning to influence the wake directly. A similar modification of vortex evolution close to a free surface was observed previously for an accelerating circular cylinder \cite{ashraf2026experimental}.

The vortex area in Fig.~\ref{fig:Av_vs_h} provides a complementary view. The general tendency is again for $A_v/D^2$ to increase as the cylinder approaches the surface, but the dependence on PPI differs from that of circulation. The 10 PPI cylinder produces the largest vortex region over much of the trajectory at $Fr=0.37$, $1.02$, and $2.00$. Its nominal-contact values are approximately $1.08$, $1.03$, and $1.11$, respectively. By comparison, the smooth-cylinder vortex area remains close to $0.6$ over these three conditions.

At the highest velocity the ordering changes. The nominal-contact vortex area for the 20 PPI cylinder reaches approximately $1.03$, larger than the values of about $0.85$, $0.70$, and $0.65$ for the 10 PPI, 60 PPI, and smooth cylinders, respectively. The 60 PPI case is particularly interesting because its vortex area decreases as $Fr$ is increased, even though, as shown below, the associated pressure deficit becomes very strong. This again shows that the influence of the porous surface cannot be represented adequately by vortex size alone.

The circulation and area results together suggest that the porous cylinders alter both the amount of rotational motion carried by the wake and the way in which that motion is distributed spatially. For example, a large-area wake such as that of the 10 PPI cylinder does not always have the most negative pressure, while the smaller vortex region of the 60 PPI cylinder can sustain a considerably stronger pressure deficit. This distinction is important because the free-surface response is ultimately affected by the pressure and momentum distribution produced by the wake rather than by a geometric measure of vortex size alone.

\subsection{Vortex-associated pressure field}

Figure~\ref{fig:Cpv_vs_h} shows the mean pressure coefficient calculated within the identified vortex region. In contrast with the vortex area, for which 10 PPI is frequently the largest case, the strongest mean pressure deficit is generally associated with the 60 PPI cylinder once the Froude number is increased above the lowest condition.

At $Fr=0.37$, the three porous cylinders develop broadly
comparable negative values of $\overline{C}_{p}^{\,v}$ during the
submerged part of the motion, while the smooth cylinder remains much
closer to zero over a substantial part of the trajectory. At nominal
contact, $\overline{C}_{p}^{\,v}$ is approximately $-0.28$, $-0.32$,
and $-0.22$ for the 10, 20, and 60 PPI cylinders, respectively,
whereas the value for the smooth cylinder is close to zero.

The behaviour changes markedly at the intermediate Froude numbers. At $Fr=1.02$, the value at the nominal contact position for 60 PPI decreases to approximately $-0.44$, compared with $-0.35$, $-0.27$, and $-0.11$ for the 20 PPI, 10 PPI, and smooth cases. An even stronger difference is observed at $Fr=2.00$, where the 60 PPI value is approximately $-0.48$. The other three cases are substantially less negative. At $Fr=4.08$, the magnitude of the deficit decreases for most of the surfaces, but 60 PPI still retains the strongest negative pressure, with $\overline{C}_{p}^{\,v}\simeq-0.34$ at the nominal contact position, $h=0.5$.

These results are useful because they separate two effects that would otherwise be difficult to distinguish from the vorticity fields alone. The 10 PPI cylinder often generates the largest vortex region and, at low and intermediate $Fr$, one of the largest circulations. The 60 PPI cylinder, however, frequently produces the lowest mean pressure inside the identified vortex. In other words, the strongest pressure signature is not obtained simply by producing the largest vortex.

A more complete measure is obtained by integrating the pressure deficit over the vortex region. The resulting quantity, $I_{p,v}$, is shown in Fig.~\ref{fig:Ipv_vs_h}. This quantity includes both the intensity of the pressure deficit and the spatial extent over which it acts. Its evolution therefore differs from that of either $A_v/D^2$ or $\overline{C}_{p}^{\,v}$ considered separately.

At $Fr=0.37$, $I_{p,v}$ grows steadily for the porous cylinders during the approach to the surface. The values at the nominal contact position for 10 and 20 PPI are nearly identical, approximately $0.307$ and $0.310$, respectively, while the 60 PPI value is lower at about $0.196$. The smooth cylinder is clearly separated from all three, with $I_{p,v}\simeq0.020$.

At $Fr=1.02$, the ranking changes. The 60 PPI cylinder now produces the largest integrated pressure deficit, approximately $0.370$ at the nominal contact position, $h=0.5$. The corresponding values for 20 PPI and 10 PPI are about $0.303$ and $0.284$, while the smooth case remains considerably lower at approximately $0.083$. The difference becomes even clearer at $Fr=2.00$: $I_{p,v}$ for 60 PPI is about $0.352$, compared with approximately $0.20$ for both 10 and 20 PPI and $0.164$ for the smooth cylinder. At the highest Froude number, the absolute values decrease, but the same ordering largely remains: 60 PPI gives the largest value at the nominal contact position ($\simeq0.243$), followed by 20 PPI ($\simeq0.151$), 10 PPI ($\simeq0.077$), and the smooth cylinder ($\simeq0.053$).

The integrated-pressure result emphasizes the non-monotonic role of the porous surface. At low $Fr$, the relatively large vortex area and circulation of the 10 and 20 PPI cylinders dominate the response. At the intermediate and higher Froude numbers, the much stronger negative pressure generated by the 60 PPI case becomes more important, despite its smaller vortex area. Thus, two wakes with similar circulation need not produce the same pressure signature, and two wakes of similar size can have substantially different pressure deficits.

A possible interpretation is that flow penetration into and within
the porous layer modifies the roll-up and organization of the
separated shear layers rather than simply reducing their strength.
The present measurements do not resolve the velocity inside the porous
material, so the detailed mechanism cannot be established directly.
Nevertheless, the external PIV and reconstructed pressure fields show
clearly that the porous coating changes the balance between vortex
circulation, vortex size, and pressure within the wake.

\subsection{Wake state at the nominal contact position, $h=0.5$}

\begin{figure*}[t]
    \centering
    \includegraphics[width=0.98\textwidth]
    {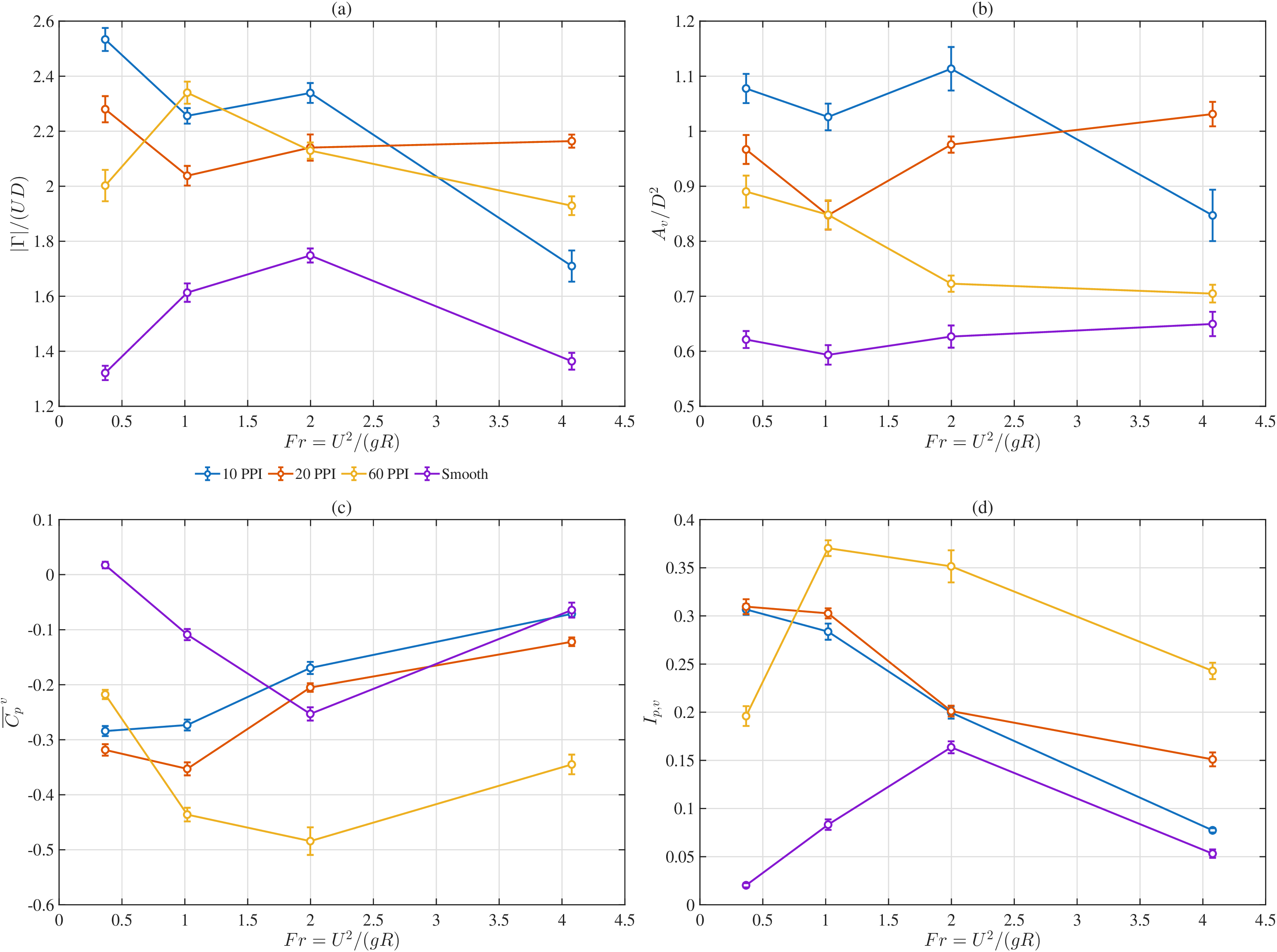}
    \caption{
    Wake properties evaluated around the nominal contact position,
    $h=0.5$, as functions of the Froude number,
    $Fr=U^2/(gR)$, for the 10 PPI, 20 PPI, 60 PPI,
    and smooth cylinders:
    (a) nondimensional circulation, $|\Gamma|/(UD)$;
    (b) nondimensional vortex area, $A_v/D^2$;
    (c) mean pressure coefficient within the vortex region,
    $\overline{C}_{p}^{\,v}$; and
    (d) integrated vortex-associated pressure deficit,
    $I_{p,v}$.
    Symbols represent averages of the instantaneous values over
    $0.45\leq h\leq0.55$, and error bars denote the corresponding
    standard deviation within this positional interval.
    Together, the four quantities characterize the strength,
    spatial extent, pressure intensity, and integrated pressure
    signature of the wake around the nominal contact position.
    }
    \label{fig:first_contact_wake}
\end{figure*}

The preceding trends are summarized in Fig.~\ref{fig:first_contact_wake}, where the wake quantities are compared at the nominal contact position, $h=0.5$. To avoid basing the comparison on a single instantaneous PIV frame, the plotted values were obtained by averaging over the narrow interval $0.45\leq h\leq0.55$; the error bars indicate the corresponding standard deviation.

Figure~\ref{fig:first_contact_wake}(a) confirms that the smooth cylinder generally produces the lowest circulation. The difference is particularly large at $Fr=0.37$, where the circulation of the 10 PPI cylinder is almost twice that of the smooth cylinder. The smooth case increases to a maximum near $Fr=2$ before decreasing again at $Fr=4.08$. The porous cases follow different trends. The circulation of 10 PPI decreases overall with increasing $Fr$, although with a local increase at $Fr=2$, whereas the 20 PPI values remain comparatively uniform after the first condition. The 60 PPI response reaches its largest value near $Fr=1$ and then decreases.

The area comparison in Fig.~\ref{fig:first_contact_wake}(b) shows an equally clear separation between the surfaces. The smooth-cylinder vortex remains relatively compact, with $A_v/D^2$ between approximately $0.59$ and $0.65$. The 10 PPI cylinder has the largest area over the first three Froude numbers, whereas 20 PPI becomes the largest at $Fr=4.08$. The 60 PPI vortex area, in contrast, decreases progressively over much of the range.

The pressure results in Figs.~\ref{fig:first_contact_wake}(c) and \ref{fig:first_contact_wake}(d) reveal a different ordering. The 60 PPI cylinder produces the most negative $\overline{C}_{p}^{\,v}$ at $Fr=1.02$, $2.00$, and $4.08$, and correspondingly gives the largest integrated pressure deficit over the same range. This is an important point: the surface producing the largest vortex is not necessarily the surface producing the strongest pressure response. The wake state at the nominal contact position, $h=0.5$, is therefore controlled by both the circulation carried by the vortex and how concentrated that rotational motion is within the pressure field.

The nominal-contact comparison also provides a useful connection with the free-surface measurements in Fig.~\ref{fig:hstar_combined}. The porous cylinders produce lower free-surface elevations than the smooth cylinder even though their submerged wakes can possess larger circulation, larger vortex area, and, in several cases, a substantially stronger vortex-associated pressure deficit. The lower surface elevation of the porous cases therefore cannot be interpreted simply as evidence of a weaker wake. Rather, the porous coating changes how the flow and pressure are distributed around the body and within the porous surface layer. The interface response and the wake metrics capture different aspects of this redistribution.

\subsection{Relation between circulation and the integrated pressure deficit}

\begin{figure*}[t]
    \centering
    \includegraphics[width=0.98\textwidth]
    {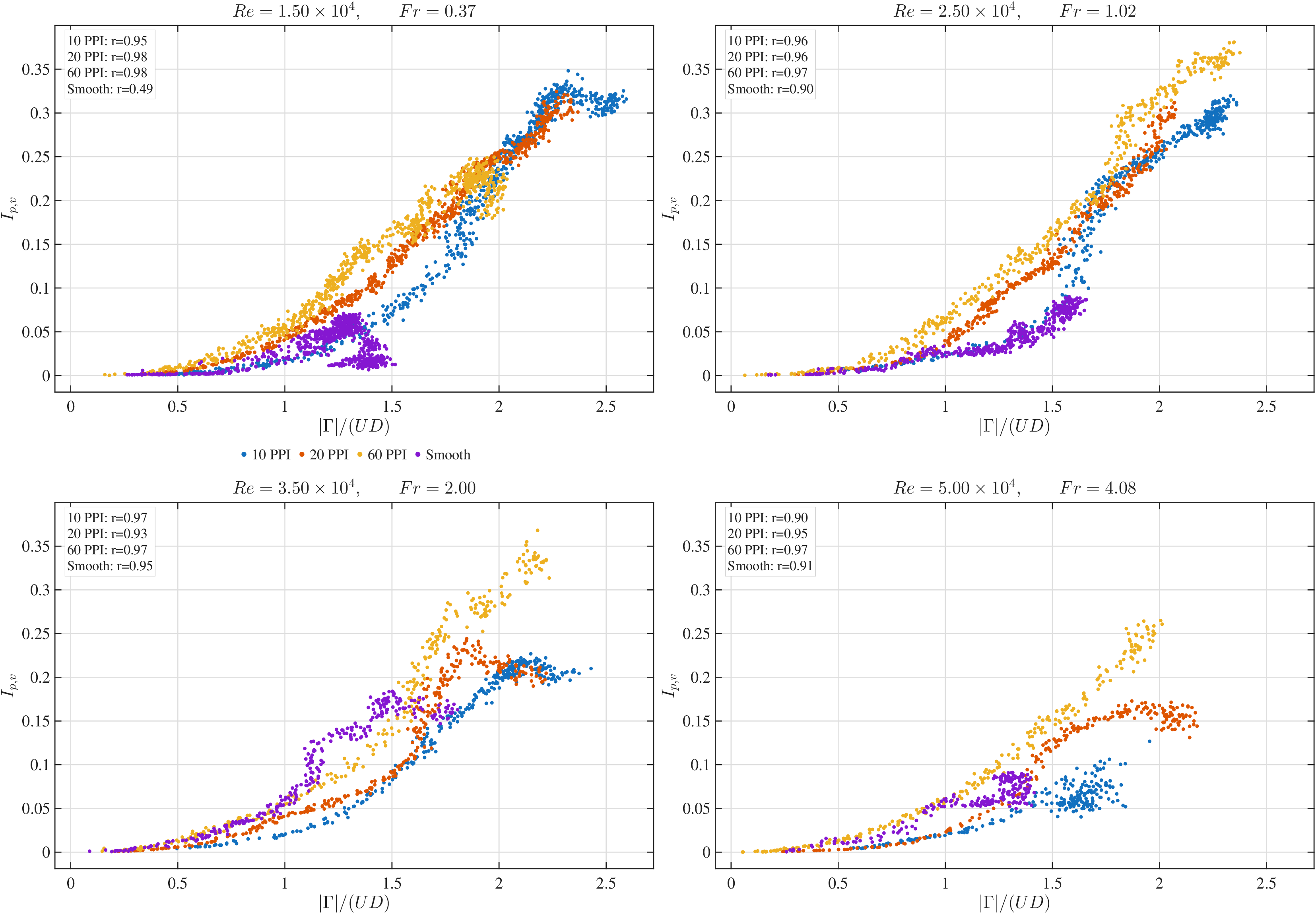}
    \caption{ Relationship between the integrated vortex-associated pressure deficit, $I_{p,v}$, and nondimensional circulation, $|\Gamma|/(UD)$, during the submerged approach to the nominal contact position ($h\geq0.5$). Results are shown for the 10 PPI, 20 PPI, 60 PPI, and smooth cylinders at $(Re,Fr)=(1.50\times10^{4},0.37)$, $(2.50\times10^{4},1.02)$, $(3.50\times10^{4},2.00)$, and $(5.00\times10^{4},4.08)$. Each point represents an instantaneous wake state prior to the nominal contact position. The Pearson correlation coefficient $r$ is reported in each panel. For the 10 PPI, 20 PPI, 60 PPI, and smooth cylinders, respectively, the numbers of instantaneous PIV states used in the correlations are $n=(705,710,693,742)$ at $Fr=0.37$, $n=(433,433,440,433)$ at $Fr=1.02$, $n=(306,307,295,301)$ at $Fr=2.00$, and $n=(223,222,234,216)$ at $Fr=4.08$. These values represent temporally sampled PIV states from a single experimental realization for each condition and not statistically independent experimental repeats. The predominantly high Pearson coefficients indicate strong co-evolution of the integrated pressure deficit and vortex circulation during the submerged approach.}
    \label{fig:Ipv_vs_Gamma}
\end{figure*}

The relationship between circulation and pressure is examined directly in Fig.~\ref{fig:Ipv_vs_Gamma}, where $I_{p,v}$ is plotted against $|\Gamma|/(UD)$ using all of the instantaneous data obtained before the nominal contact position ($h\geq0.5$). A strong positive relationship is evident for almost all of the porous-cylinder cases.

At $Fr=0.37$, the correlation coefficients for the 10, 20, and 60 PPI cylinders are $r=0.95$, $0.98$, and $0.98$, respectively. The smooth cylinder behaves differently at this condition, with a much weaker correlation of $r=0.49$. At $Fr=1.02$, strong correlations are recovered for all four surfaces, with $r$ between $0.90$ and $0.97$. Similarly, at $Fr=2.00$ the values range from $0.93$ to $0.97$, and at $Fr=4.08$ they remain between $0.90$ and $0.97$.

The high correlation coefficients show that the integrated pressure
deficit and circulation co-evolve strongly during the submerged
approach to the free surface. This behaviour is particularly clear for
the porous cylinders, for which the two quantities follow well-defined
trajectories over most of the approach. However, both quantities also
vary systematically with cylinder position $h$, and successive
instantaneous PIV samples are temporally correlated. The Pearson
coefficients are therefore interpreted here as descriptive measures of
their co-evolution rather than as evidence of statistical independence
or a causal relation between the two quantities. Within the present
operating conditions, circulation nevertheless provides a useful global
indicator of the evolving pressure-bearing strength of the wake.

The relationship is not, however, a single universal curve. At a given value of $|\Gamma|/(UD)$, different PPI cases can have noticeably different $I_{p,v}$. This is most evident for the 60 PPI cylinder at the intermediate and high Froude numbers, where the integrated pressure deficit is larger than that of the other cases at comparable circulation. The differences are consistent with the vortex-area and mean-pressure results: circulation describes the overall rotational strength of the wake, but the resulting pressure deficit also depends on how that circulation is distributed spatially.

The present results therefore point to a more complicated role of the porous-surface configuration than simple wake suppression. The 10 PPI surface tends to produce a relatively large vortex with high circulation, particularly at low and intermediate $Fr$. The 60 PPI surface often produces a smaller vortex but a much stronger negative pressure within it, while the 20 PPI case becomes particularly persistent at the highest Froude number. The smooth cylinder, by comparison, generally has a smaller vortex region and lower integrated pressure deficit. The porous coating thus changes the balance between vortex strength,
size, and pressure rather than shifting all of these quantities in the
same direction.

\begin{figure*}[t]
    \centering

    \begin{subfigure}[t]{0.9\textwidth}
        \centering
        \includegraphics[width=\textwidth]{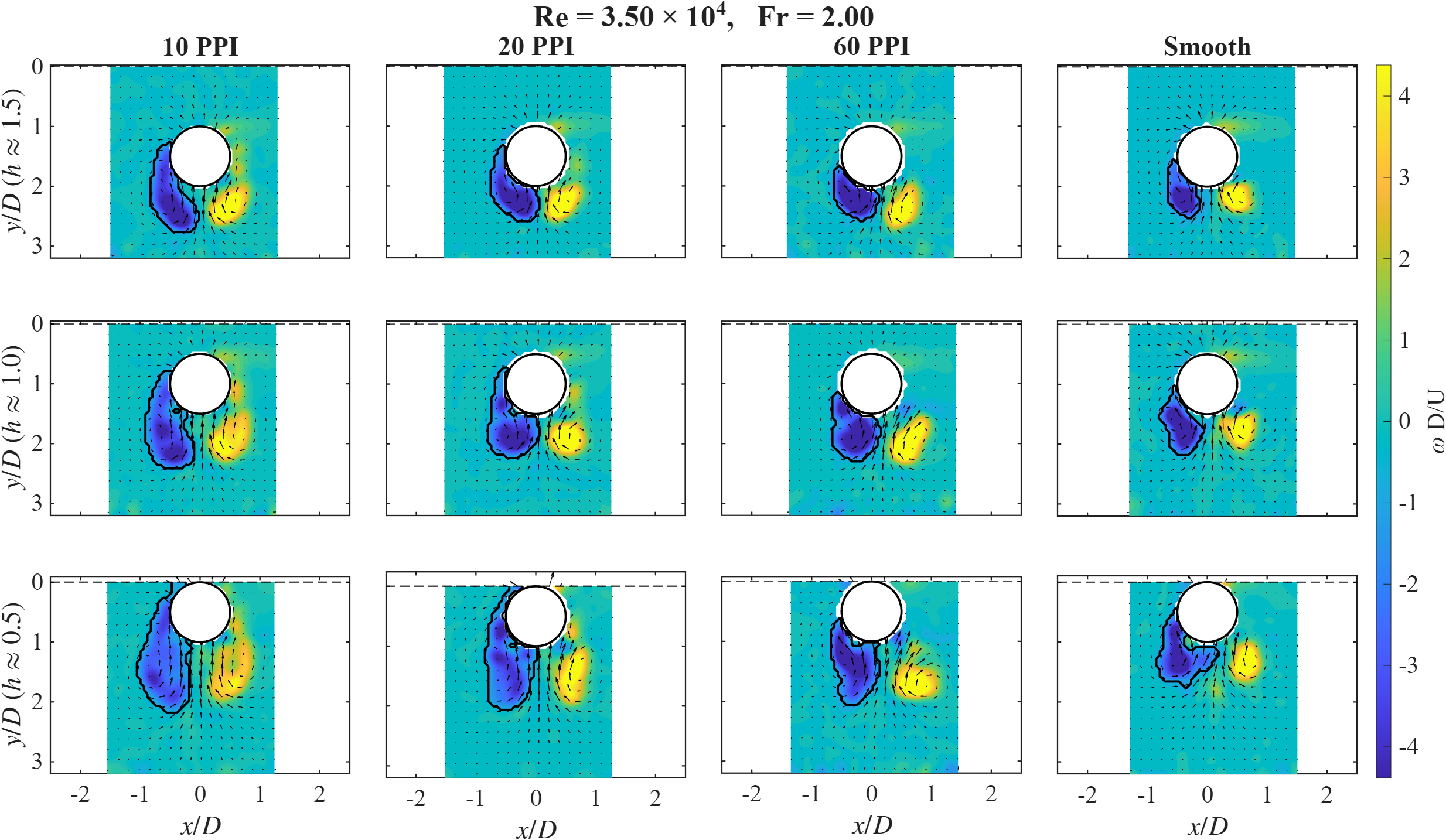}
        \caption{}
        \label{fig:flow_vorticity_Re35000}
    \end{subfigure}

    \vspace{0.5em}

    \begin{subfigure}[t]{0.9\textwidth}
        \centering
        \includegraphics[width=\textwidth]{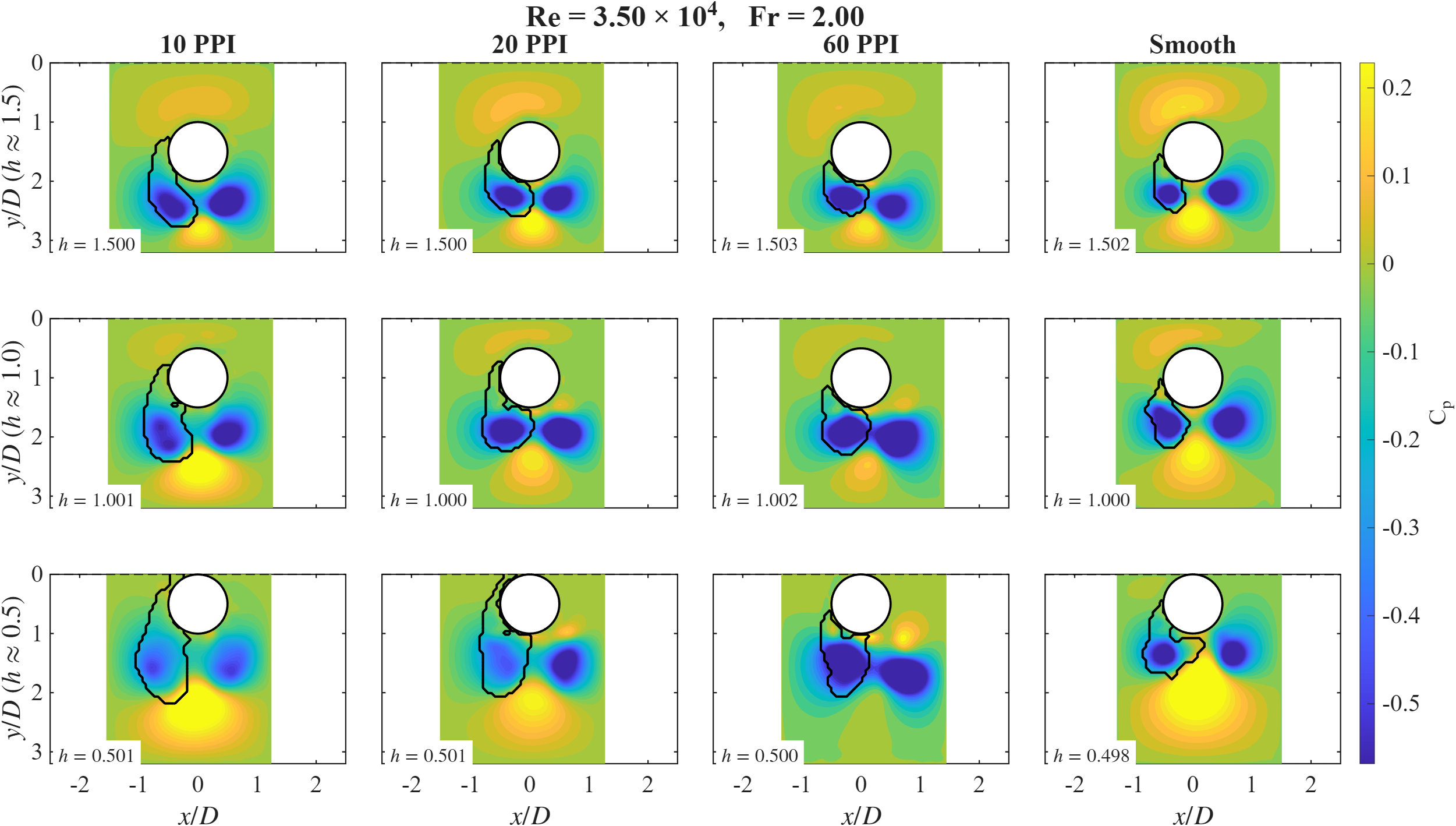}
        \caption{}
        \label{fig:flow_pressure_Re35000}
    \end{subfigure}

    \caption{
    Evolution of the wake structure and the associated pressure field during
    water exit at $Re=3.50\times10^{4}$ and $Fr=2.00$ for the 10 PPI,
    20 PPI, 60 PPI, and smooth cylinders.
    Columns correspond, from left to right, to 10 PPI, 20 PPI, 60 PPI,
    and the smooth cylinder, while the rows show three representative
    cylinder positions, $h\approx1.5$, $1.0$, and $0.5$, with
   $h=0.5$ denoting the nominal contact position, at which the upper surface of the cylinder reaches the initially undisturbed free-surface level.
    (a) Nondimensional vorticity, $\omega D/U$, with the in-plane velocity
    vectors superimposed.
    (b) Corresponding pressure-coefficient field, $C_p$.
    The black contour delineates the clockwise vortex selected for the
    quantitative analysis. Circulation, vortex area, and the
    vortex-associated pressure quantities are evaluated within this
    identified region.
    The white circular region denotes the cylinder, and the horizontal
    dashed line indicates the undisturbed free-surface level.
    The same colour scale is used for all cases within each panel to enable
    direct comparison between the porous and smooth cylinders.
    }
    \label{fig:wake_pressure_evolution_Re35000}
\end{figure*}

\begin{figure*}[t]
    \centering
    \includegraphics[width=0.98\textwidth]{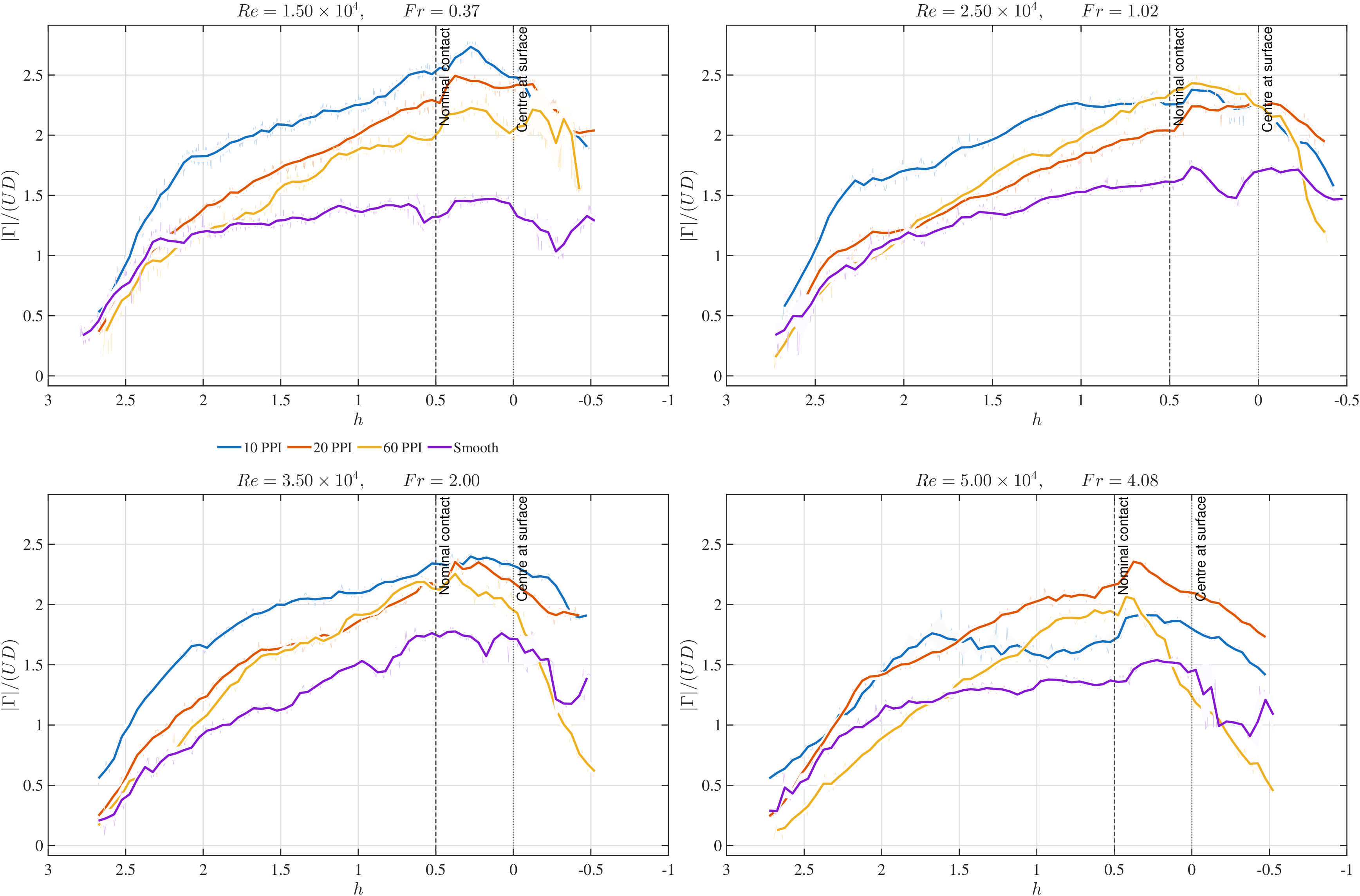}
    \caption{
    Evolution of the nondimensional vortex circulation,
    $|\Gamma|/(UD)$, as a function of the nondimensional
    cylinder position $h$ for the 10 PPI, 20 PPI, 60 PPI,
    and smooth cylinders.
    The four panels correspond to
    $(Re,Fr)=(1.50\times10^{4},0.37)$,
    $(2.50\times10^{4},1.02)$,
    $(3.50\times10^{4},2.00)$, and
    $(5.00\times10^{4},4.08)$.
    The vertical dashed line at $h=0.5$ denotes the nominal contact
    position, at which the upper surface of the cylinder reaches the
    initially undisturbed free-surface level,
    while the vertical dotted line at $h=0$ indicates the instant
    at which the cylinder centre reaches the undisturbed
    free-surface level.
    The comparison shows the systematic influence of the porous coating
    on the development and persistence of wake circulation during the
    approach to and passage through the free surface.
    }
    \label{fig:gamma_vs_h}
\end{figure*}

\begin{figure*}[t]
    \centering
    \includegraphics[width=0.98\textwidth]{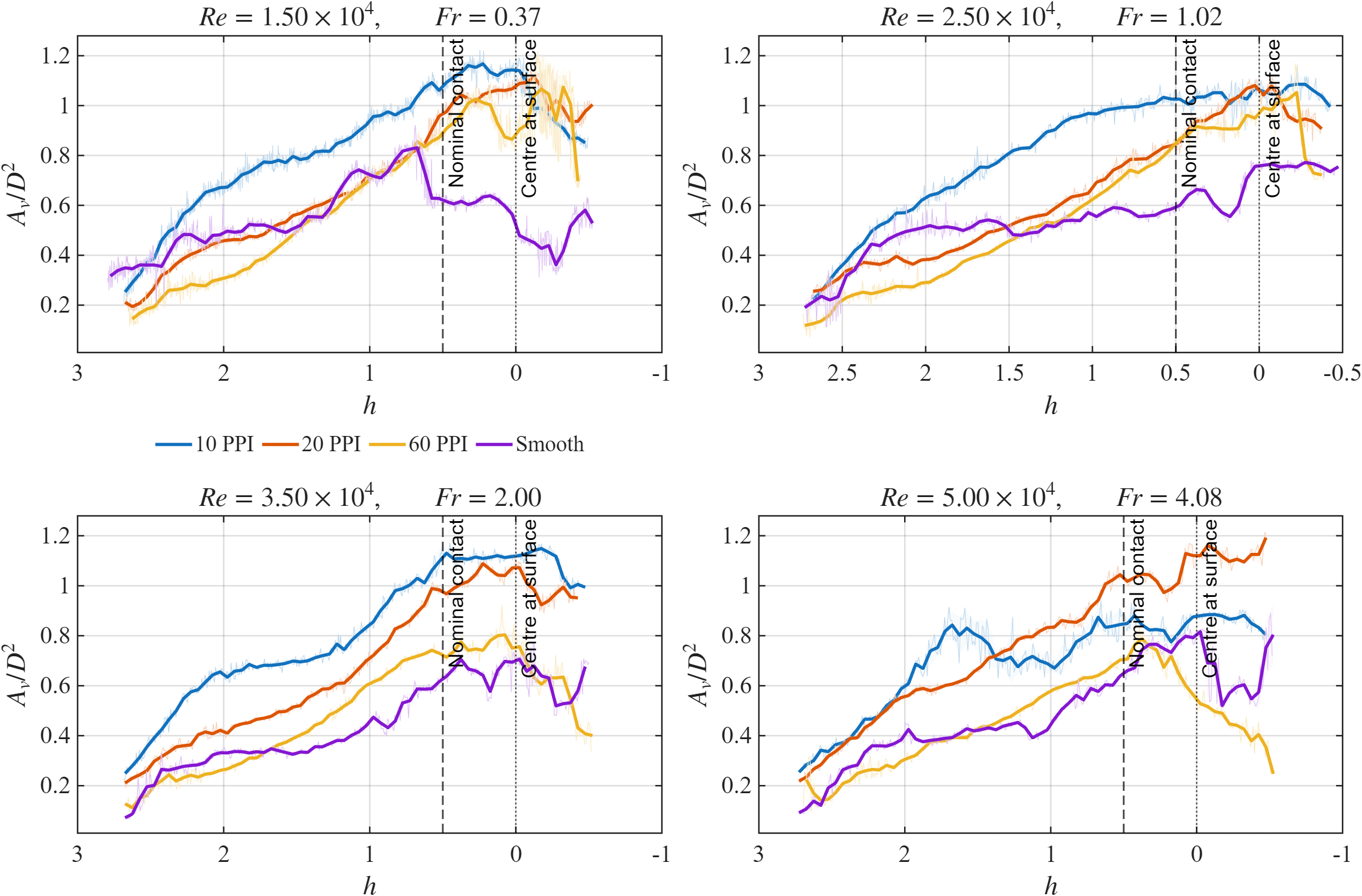}
    \caption{
    Evolution of the nondimensional vortex area,
    $A_v/D^2$, with cylinder position $h$ for the 10 PPI,
    20 PPI, 60 PPI, and smooth cylinders.
    Results are shown for
    $(Re,Fr)=(1.50\times10^{4},0.37)$,
    $(2.50\times10^{4},1.02)$,
    $(3.50\times10^{4},2.00)$, and
    $(5.00\times10^{4},4.08)$.
    The dashed and dotted vertical lines denote nominal contact
    ($h=0.5$) and passage of the cylinder centre through the
    initially undisturbed free surface ($h=0$), respectively.
    The porous cylinders generally sustain a larger vortex region
    than the smooth cylinder, although the magnitude and subsequent evolution depend strongly on both pore density and Froude number.
    }
    \label{fig:Av_vs_h}
\end{figure*}

\begin{figure*}[t]
    \centering
    \includegraphics[width=0.98\textwidth]{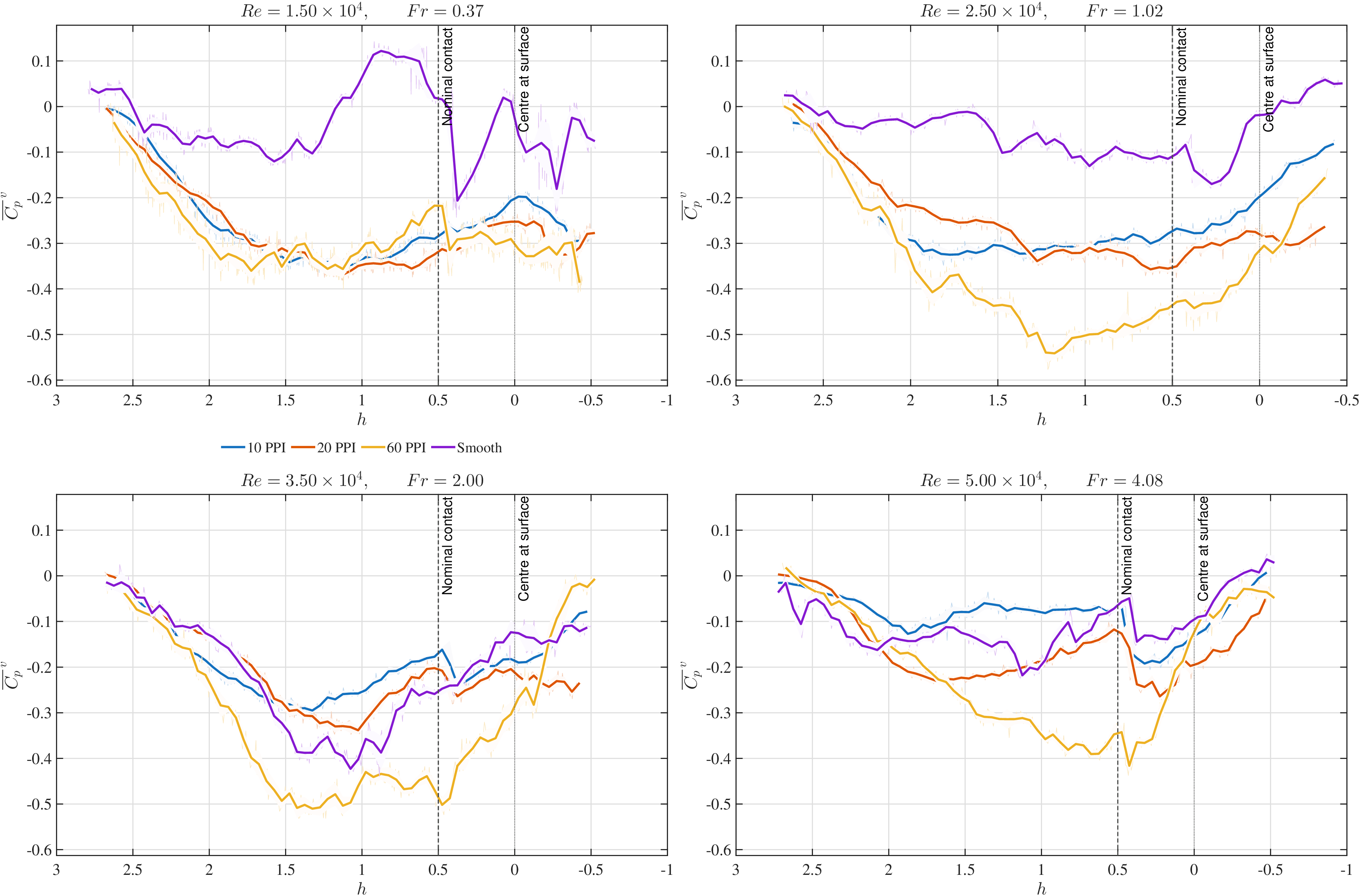}
    \caption{
    Evolution of the mean pressure coefficient within the
    identified vortex region, $\overline{C}_{p}^{\,v}$,
    as a function of cylinder position $h$.
    The four panels correspond to
    $(Re,Fr)=(1.50\times10^{4},0.37)$,
    $(2.50\times10^{4},1.02)$,
    $(3.50\times10^{4},2.00)$, and
    $(5.00\times10^{4},4.08)$.
    Results are compared for the 10 PPI, 20 PPI, 60 PPI,
    and smooth cylinders.
    More negative values of $\overline{C}_{p}^{\,v}$ represent
    a stronger mean pressure deficit within the vortex region.
    The dashed line marks the nominal contact position at $h=0.5$, whereas
    the dotted line marks $h=0$, when the cylinder centre
    reaches the initially undisturbed free surface.
    The results demonstrate that the porous coating modifies not only
    the vortex geometry and circulation but also the intensity of the
    associated low-pressure region.
    }
    \label{fig:Cpv_vs_h}
\end{figure*}

\begin{figure*}[t]
    \centering
    \includegraphics[width=0.98\textwidth]{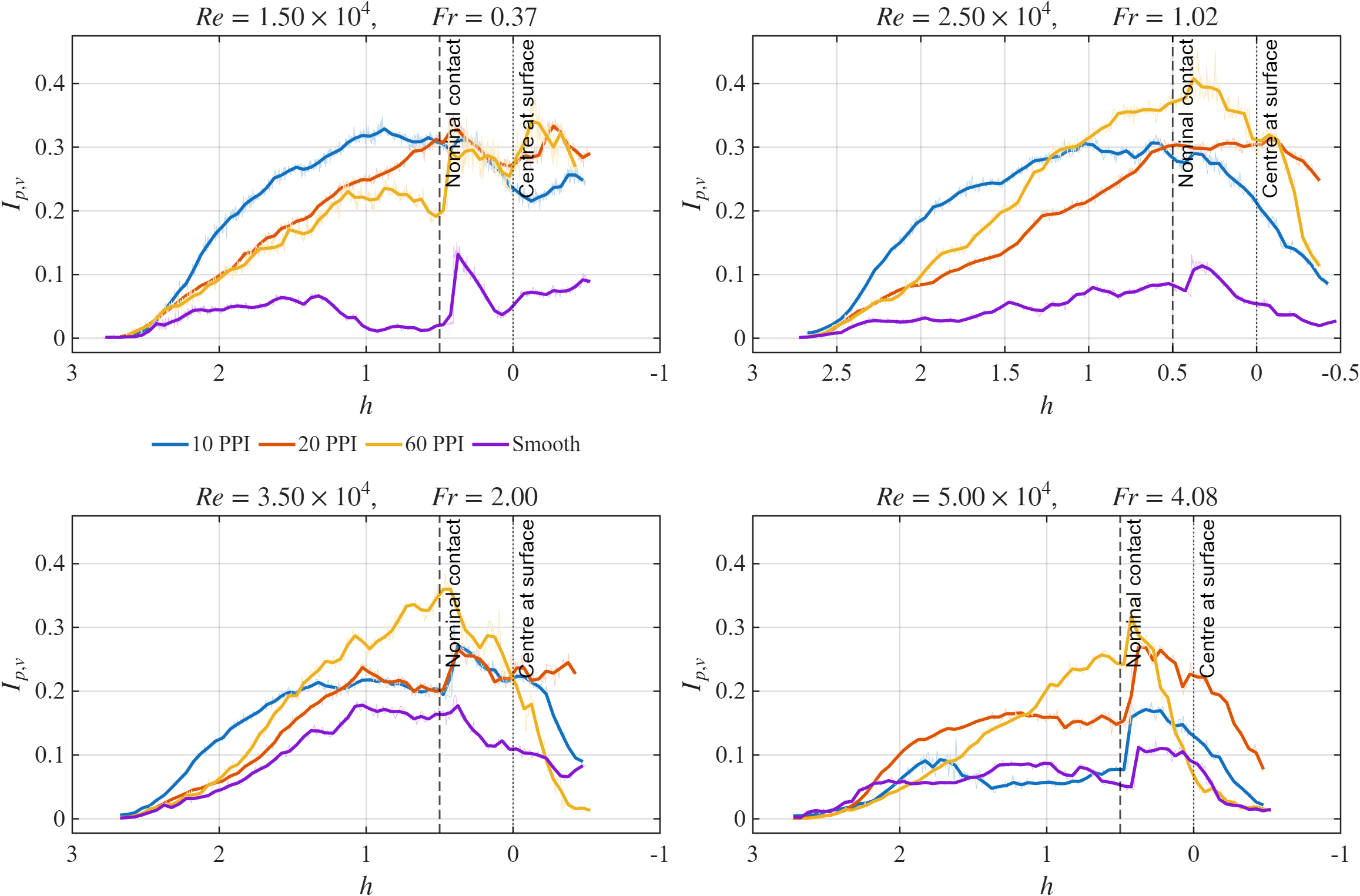}
    \caption{
    Evolution of the integrated vortex-associated pressure
    deficit, $I_{p,v}$, with nondimensional cylinder position $h$
    for the 10 PPI, 20 PPI, 60 PPI, and smooth cylinders.
    The four panels correspond to
    $(Re,Fr)=(1.50\times10^{4},0.37)$,
    $(2.50\times10^{4},1.02)$,
    $(3.50\times10^{4},2.00)$, and
    $(5.00\times10^{4},4.08)$.
    The dashed vertical line denotes the nominal contact position at $h=0.5$,
    and the dotted line denotes $h=0$.
    Because $I_{p,v}$ accounts for both the pressure deficit
    and its spatial extent within the identified vortex region,
    it provides an integrated measure of the pressure signature
    associated with the wake. The pronounced differences between the porous and smooth cylinders demonstrate the strong influence of the porous coating on the pressure-bearing wake structure.
    }
    \label{fig:Ipv_vs_h}
\end{figure*}

This distinction is important for water exit because the wake formed while the cylinder is submerged is the flow structure that subsequently encounters the free surface. The results in Figs.~\ref{fig:hstar_combined}--\ref{fig:Ipv_vs_Gamma} show that this wake has already been substantially modified by the porous surface before the nominal contact position occurs. The subsequent free-surface interaction therefore begins from a different hydrodynamic state for each surface, which provides a basis for the differences observed in the interface deformation during water exit.

\subsection{Modal organization of the wake} 
\begin{figure*}[t]
    \centering
    \includegraphics[width=0.90\textwidth]
    {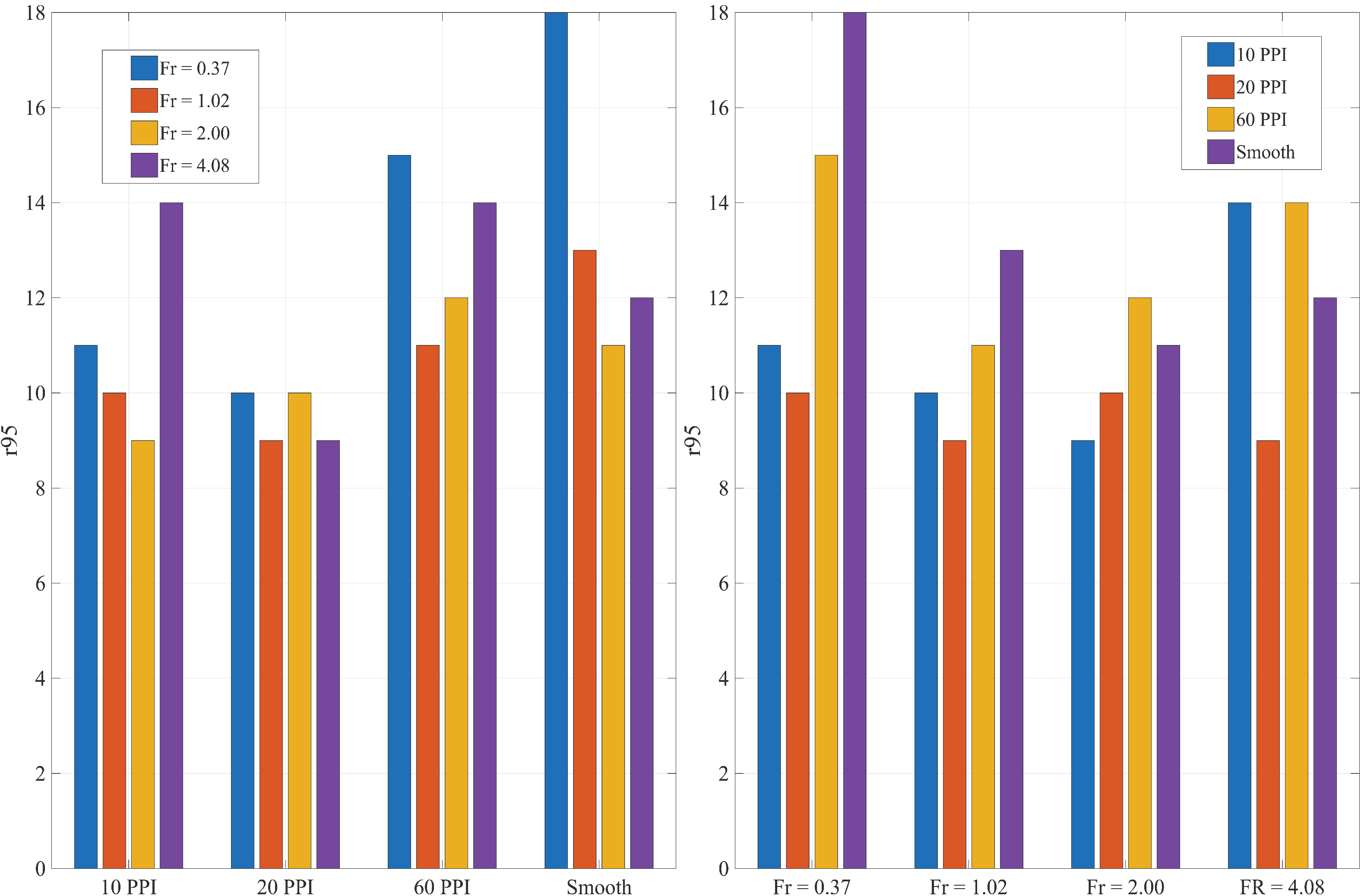}
    \caption{
    Number of POD modes, $r_{95}$, required to recover 95\% of the
    total modal energy for the 10 PPI, 20 PPI, 60 PPI, and smooth
    cylinders at the four tested operating conditions. The relatively
small values of $r_{95}$ indicate that the measured wake dynamics
are dominated by a limited number of energetic modes. The 20 PPI
configuration exhibits the most consistently compact modal
representation, requiring $r_{95}=9$--$10$ across the tested
conditions.
}
   
    \label{fig:pod_r95}
\end{figure*}

The preceding analysis characterizes a selected coherent wake vortex through its circulation, spatial extent, and pressure signature. POD provides a complementary measure of how the energetic content of the entire measured velocity field is distributed among coherent structures. Figure~\ref{fig:pod_r95} therefore compares the number of POD modes required to recover 95\% of the total modal energy for each surface and operating condition. All of the wakes are strongly dominated by a relatively small number of energetic modes. Across the four operating conditions, 95\% of the POD energy is recovered using 11, 10, 9, and 14 modes for the 10 PPI cylinder; 10, 9, 10, and 9 modes for the 20 PPI cylinder; 15, 11, 12, and 14 modes for the 60 PPI cylinder; and 18, 13, 11, and 12 modes for the smooth cylinder, respectively. Thus, only approximately 9--18 modes are required to represent 95\% of the measured fluctuation energy despite the considerably larger number of available modes. A notable feature is the comparatively compact modal representation of the 20 PPI wake. Its value of $r_{95}$ remains between 9 and 10 over all four operating conditions, whereas the 10 PPI, 60 PPI, and smooth-cylinder cases exhibit larger variations and generally require more modes. Averaged across the tested conditions, $r_{95}$ is 9.5 for 20 PPI, compared with 11.0 for 10 PPI, 13.0 for 60 PPI, and 13.5 for the smooth cylinder. The dependence of modal dimensionality on surface condition is therefore also non-monotonic with pore density. This modal result complements the vortex-based quantities discussed above. The porous coating modifies not only the circulation, spatial extent, and pressure signature of the identified vortex, but also the distribution of energy among the dominant structures of the broader measured wake. In particular, the intermediate 20 PPI configuration produces the most compact energetic representation, even though it does not systematically correspond to either the weakest or strongest vortex according to the local wake measures. The POD analysis therefore provides additional evidence that the effect of the porous coating is a reorganization of the wake rather than a monotonic suppression of its strength.

% ============================================================
% FIGURE: VORTEX AREA
% ============================================================

% ============================================================
% FIGURE: MEAN PRESSURE COEFFICIENT INSIDE VORTEX
% ============================================================

% ============================================================
% FIGURE: INTEGRATED VORTEX-ASSOCIATED PRESSURE DEFICIT
% ============================================================

% ============================================================
% FIGURE: FIRST-CONTACT WAKE QUANTITIES
% ============================================================
% ============================================================
% PREFERRED VERSION OF THE FIRST-CONTACT FIGURE
% ============================================================

% ============================================================
% FIGURE: INTEGRATED PRESSURE DEFICIT VS CIRCULATION
% ============================================================

\section{\label{sec:conclusion}Conclusion}

The water-exit dynamics of a circular cylinder covered by a porous
foam layer were investigated experimentally and compared with those
of an equivalent smooth cylinder. Three porous coatings with pore
densities of 10, 20, and 60 PPI were considered over
$0.37\leq Fr\leq4.08$, corresponding to
$1.50\times10^{4}\leq Re\leq5.00\times10^{4}$.
Time-resolved PIV was used to follow the development of the submerged
wake and to reconstruct the associated pressure field as the cylinder
approached and crossed the free surface. Proper orthogonal
decomposition (POD) was additionally used to characterize the
energetic organization of the measured wake.

The porous coating produced a clear modification of both the
free-surface response and the submerged wake. All three porous
cylinders generated a lower free-surface elevation than the smooth
cylinder, with the largest reduction generally observed for the
60 PPI configuration. This reduction, however, was not accompanied
by a systematically weaker wake. Before the nominal contact position,
the porous cylinders generally developed larger vortex circulation
and vortex area than the smooth reference. The response was also
strongly non-monotonic with porous-surface configuration: the
configuration producing the largest circulation or vortex area
changed across the tested $Re$--$Fr$ conditions.

The reconstructed pressure fields revealed an additional distinction
between vortex size and vortex intensity. Although the 10 PPI cylinder
often produced the largest vortex region, the 60 PPI cylinder
developed the strongest mean negative pressure within the identified
vortex at the intermediate and higher tested conditions. Consequently,
the 60 PPI case also produced the largest integrated
vortex-associated pressure deficit, $I_{p,v}$, at the nominal contact
position, $h=0.5$, for $Fr=1.02$, $2.00$, and $4.08$. Thus, the
vortex with the largest spatial extent is not necessarily the vortex
with the strongest pressure signature.

A strong relationship was found between the integrated pressure
deficit and circulation during the submerged approach to the free
surface. For the porous cylinders, the Pearson correlation
coefficients were typically in the range $r=0.90$--$0.98$,
indicating strong co-evolution between the accumulation of circulation
and the development of the pressure-bearing wake. The data
nevertheless do not collapse onto a single universal relation, showing
that circulation alone does not determine the pressure response and
that the spatial organization of the vortex also remains important.

The POD analysis provides a complementary view of this wake
reorganization. Across all tested conditions, only 9--18 modes were
required to recover 95\% of the total modal energy, indicating that
the measured wake dynamics are dominated by a relatively small number
of energetic structures. The 20 PPI configuration showed the most
consistently compact modal representation, requiring only 9--10 modes
across the four operating conditions, compared with 9--14 modes for
10 PPI, 11--15 modes for 60 PPI, and 11--18 modes for the smooth
cylinder. The dependence of modal dimensionality on porous-surface
configuration is therefore also non-monotonic. This result shows that
the porous coating modifies not only the circulation, spatial extent,
and pressure signature of the identified vortex, but also the
distribution of energy among the dominant structures of the broader
measured wake.

Taken together, the results demonstrate that a porous coating does
not simply suppress the wake. Instead, it reorganizes the balance
between vortex circulation, spatial extent, pressure, and the
distribution of energy among the dominant wake structures, while
simultaneously reducing the free-surface elevation. The response is
distinctly non-monotonic with porous-surface configuration: the
10 PPI cylinder frequently produces the largest vortex region, the
60 PPI cylinder develops the strongest pressure deficit at the
intermediate and higher tested conditions, and the 20 PPI wake
exhibits the most compact POD representation. The submerged wake
therefore reaches the free surface in a different hydrodynamic state
for each porous configuration. Since the present measurements do not
resolve the velocity or pressure within the porous foam layer, the
internal transport responsible for these differences cannot be
established directly. Resolving the flow within the porous material,
together with independent measurements of its porosity and
permeability, would therefore be a useful next step toward identifying
the mechanism responsible for the observed dependence on
porous-surface configuration.

\section*{Acknowledgements}

The financial support of the Belgian Fund for Scientific Research under research project WOLFLOW (F.R.S.-FNRS, PDR T.0021.18) is gratefully acknowledged. Part of the experimental setup was financed by {\it Fonds Sp\'eciaux} from ULi\`ege. SD is F.R.S--FNRS Senior Research Associate.

\section*{Authors’ Contributions}

IA conceived and designed the study, performed the experiments, processed the PIV data, developed the vortex identification framework, carried out the analysis, and wrote the manuscript.  

SD supervised the research, provided guidance on experimental design and interpretation, and contributed to refining the discussion and conclusions.  

NT developed the PIV based pressure-estimation software, while MT assisted in generating the pressure data and provided technical input during the analysis.

\section*{Data Availability Statement}
The data that support the findings of this study are available from
the corresponding author upon reasonable request.

\appendix

% ============================================================
% APPENDIX
% ============================================================
\setcounter{figure}{0}
\renewcommand{\thefigure}{A\arabic{figure}}
\section{Additional wake and pressure fields}
\label{app:wake_fields}

\begin{figure*}[t]
    \centering

    \begin{subfigure}[t]{0.90\textwidth}
        \centering
        \includegraphics[width=\textwidth]
        {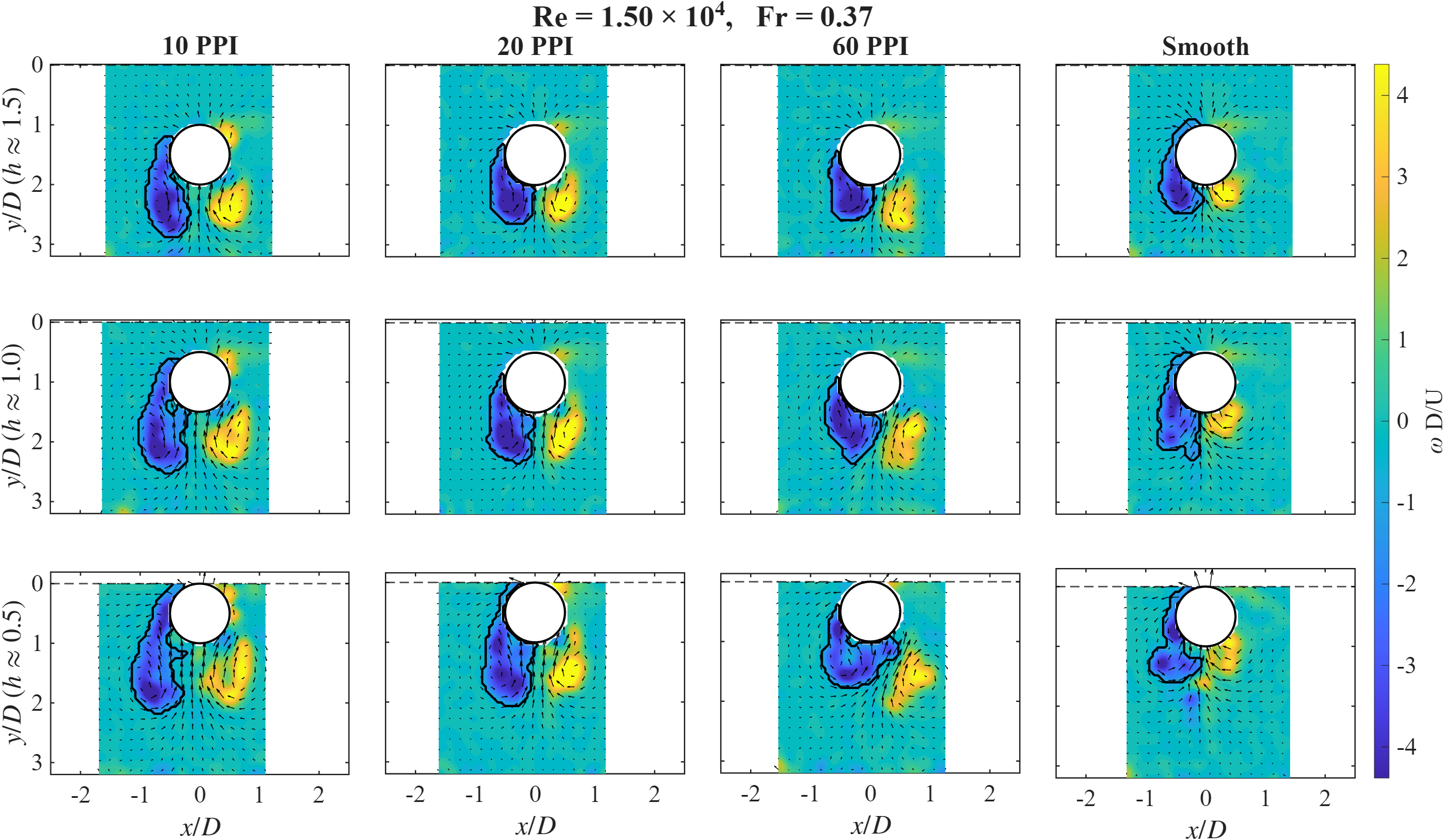}
        \caption{}
        \label{fig:appendix_vorticity_U300}
    \end{subfigure}

    \vspace{0.5em}

    \begin{subfigure}[t]{0.90\textwidth}
        \centering
        \includegraphics[width=\textwidth]
        {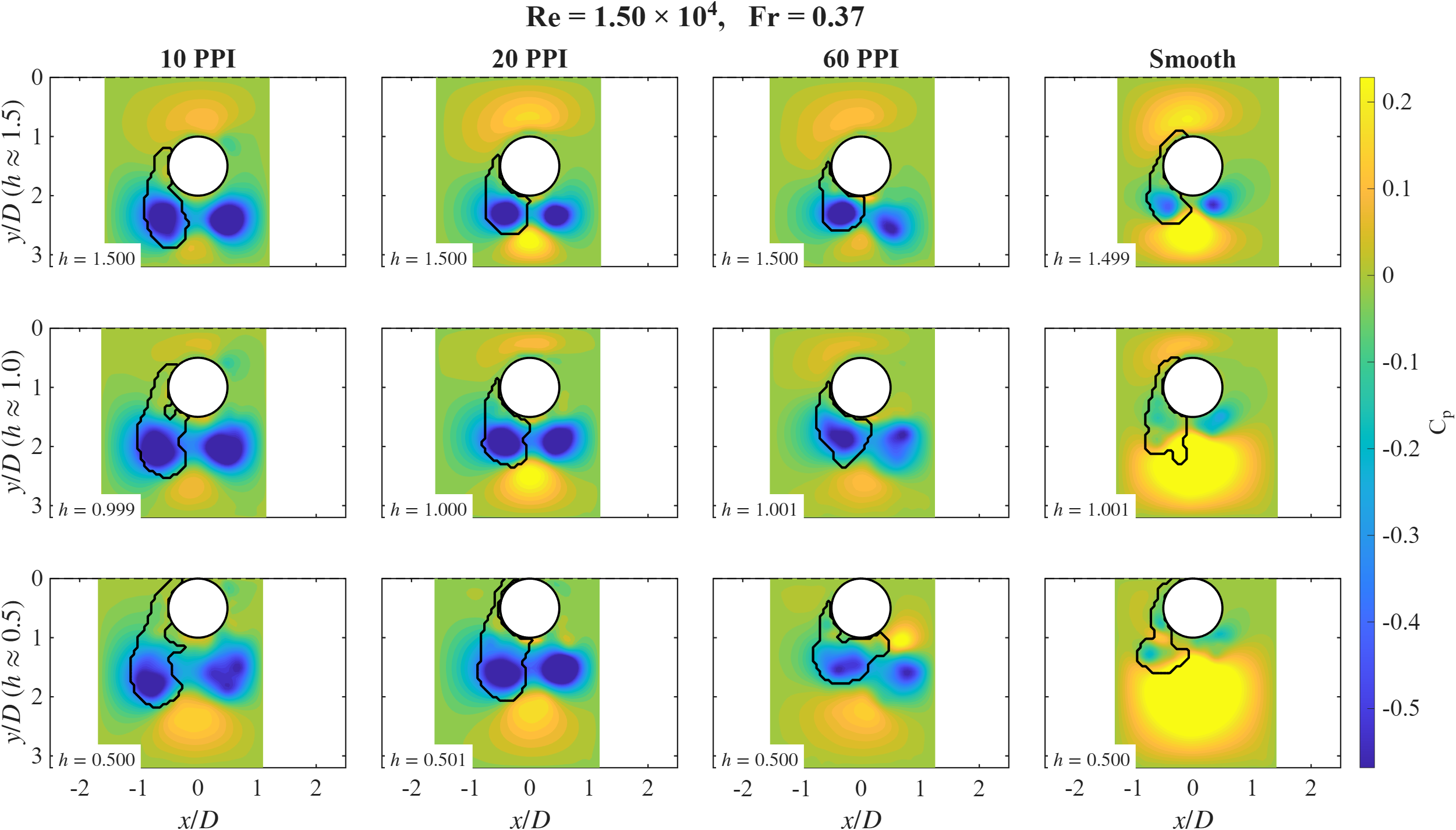}
        \caption{}
        \label{fig:appendix_cp_U300}
    \end{subfigure}

    \caption{
    Evolution of the wake and associated pressure field at
    $Re=1.50\times10^{4}$ and $Fr=0.37$ for the 10 PPI,
    20 PPI, 60 PPI, and smooth cylinders.
    Columns correspond, from left to right, to 10 PPI, 20 PPI,
    60 PPI, and the smooth cylinder, while the rows correspond to
    approximately $h=1.5$, $1.0$, and $0.5$.
    (a) Nondimensional vorticity, $\omega D/U$, with the in-plane
    velocity vectors superimposed.
    (b) Corresponding reconstructed pressure-coefficient field,
    $C_p$.
    The black contour delineates the clockwise vortex used for the
    quantitative analysis. The white circular region denotes the
    cylinder, and the horizontal dashed line indicates the initially
    undisturbed free-surface level.
    The same colour scale is used for all four surface conditions
    within each panel.
    }
    \label{fig:appendix_U300}
\end{figure*}

\begin{figure*}[t]
    \centering

    \begin{subfigure}[t]{0.90\textwidth}
        \centering
        \includegraphics[width=\textwidth]
        {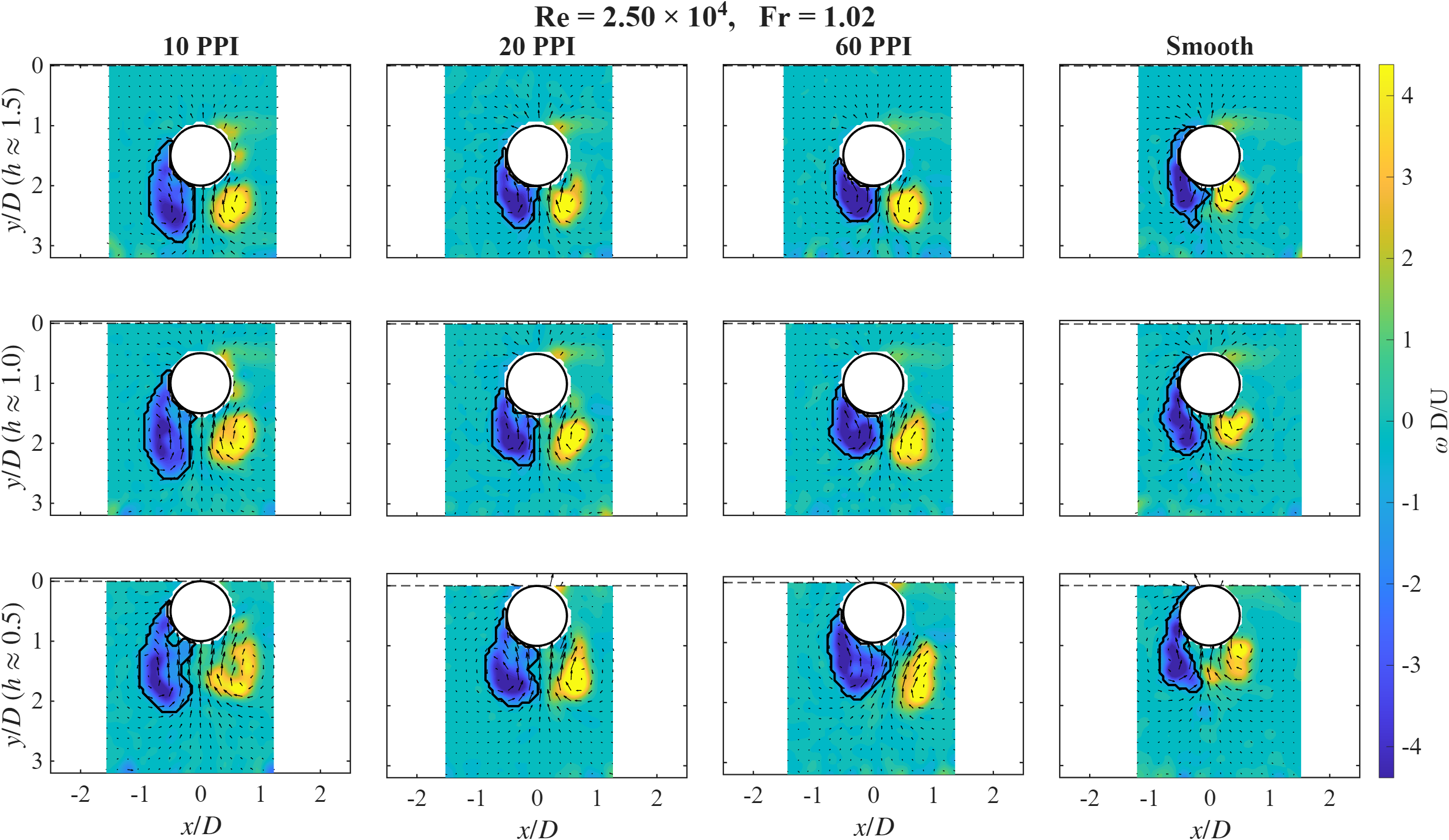}
        \caption{}
        \label{fig:appendix_vorticity_U500}
    \end{subfigure}

    \vspace{0.5em}

    \begin{subfigure}[t]{0.90\textwidth}
        \centering
        \includegraphics[width=\textwidth]
        {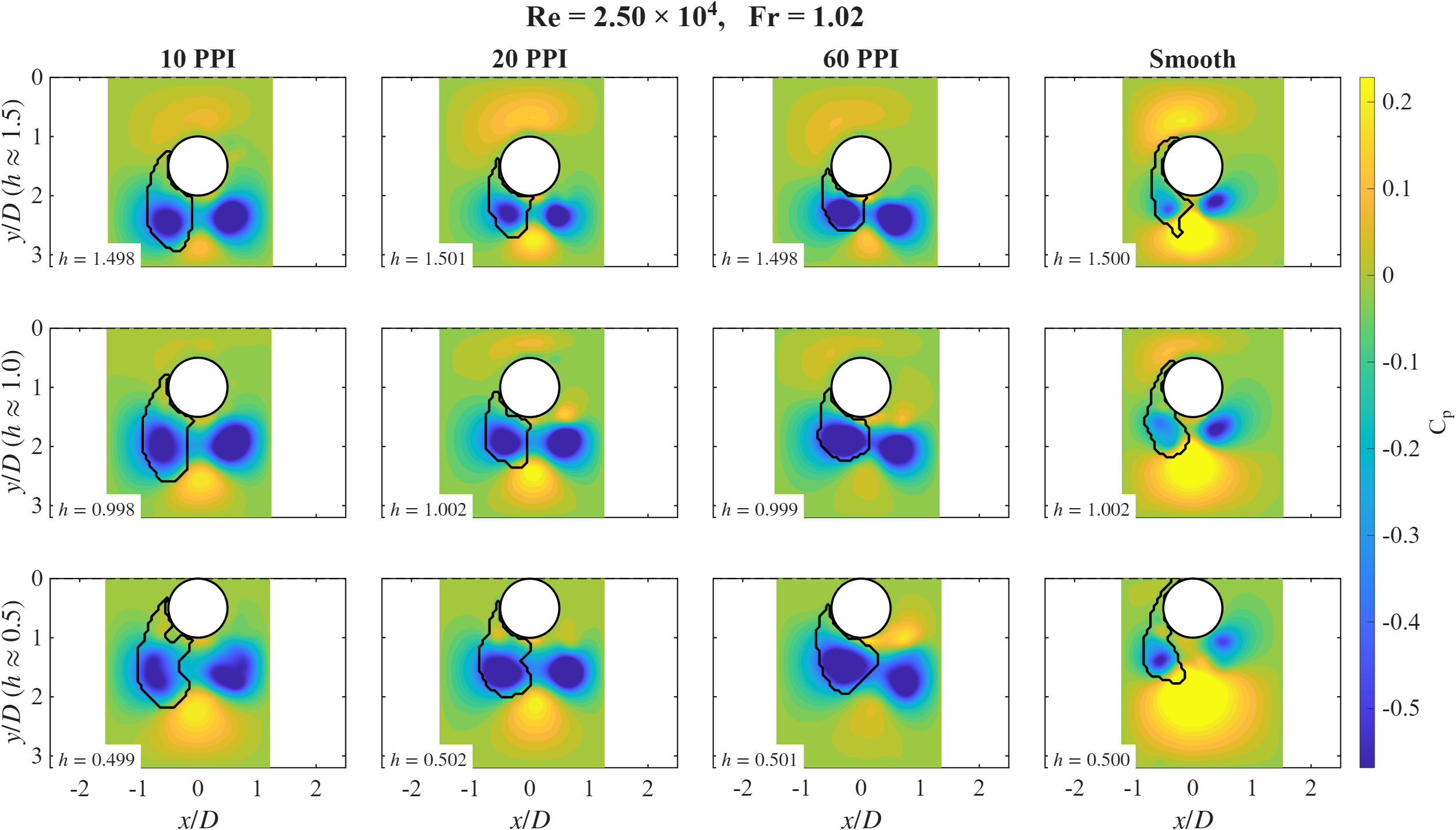}
        \caption{}
        \label{fig:appendix_cp_U500}
    \end{subfigure}

    \caption{
    Evolution of the wake and associated pressure field at
    $Re=2.50\times10^{4}$ and $Fr=1.02$ for the 10 PPI,
    20 PPI, 60 PPI, and smooth cylinders.
    Columns correspond, from left to right, to 10 PPI, 20 PPI,
    60 PPI, and the smooth cylinder, while the rows correspond to
    approximately $h=1.5$, $1.0$, and $0.5$.
    (a) Nondimensional vorticity, $\omega D/U$, with the in-plane
    velocity vectors superimposed.
    (b) Corresponding reconstructed pressure-coefficient field,
    $C_p$.
    The black contour delineates the clockwise vortex used for the
    quantitative analysis. The white circular region denotes the
    cylinder, and the horizontal dashed line indicates the initially
    undisturbed free-surface level.
    The same colour scale is used for all four surface conditions
    within each panel.
    }
    \label{fig:appendix_U500}
\end{figure*}

\begin{figure*}[t]
    \centering

    \begin{subfigure}[t]{0.90\textwidth}
        \centering
        \includegraphics[width=\textwidth]
        {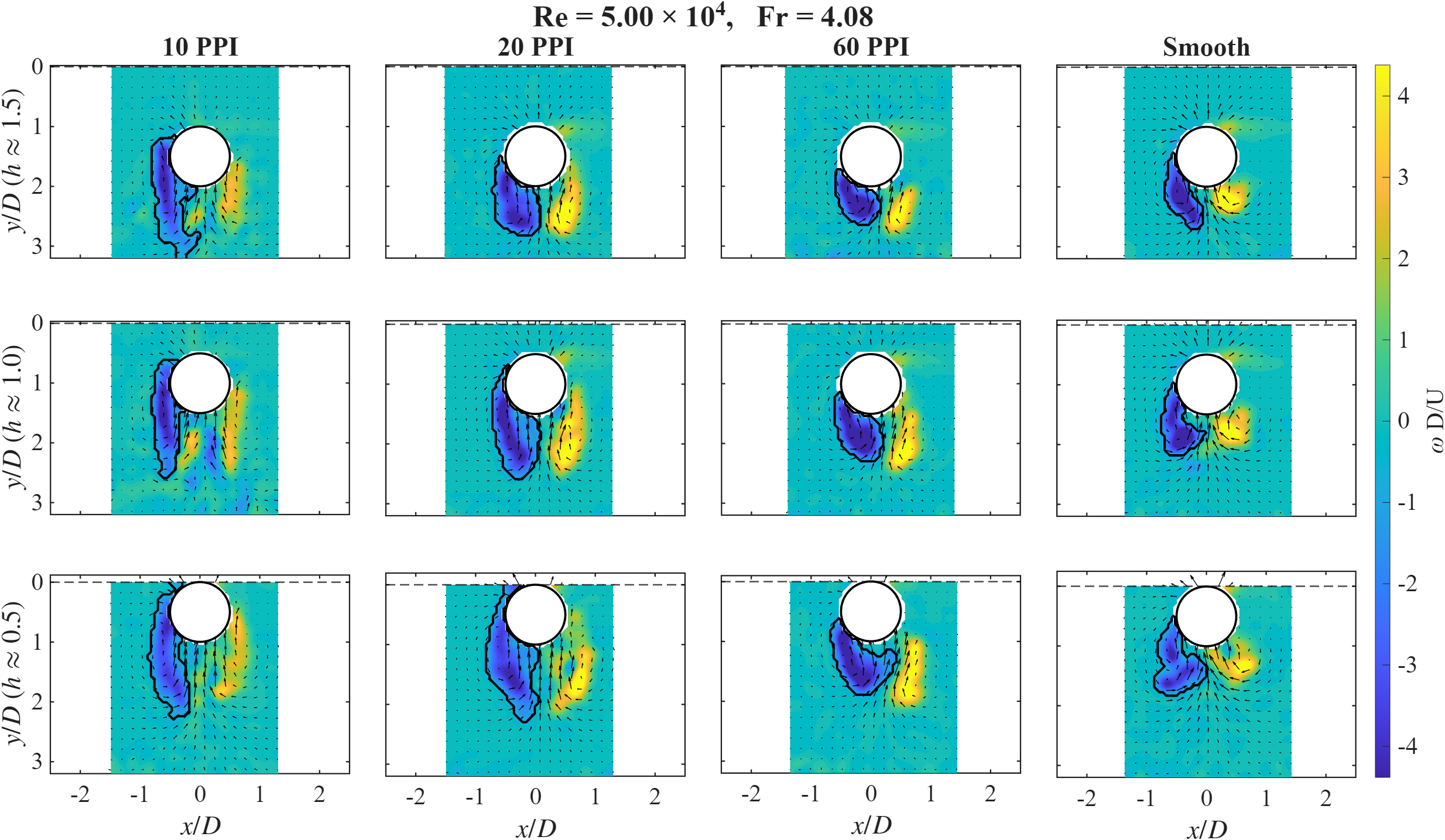}
        \caption{}
        \label{fig:appendix_vorticity_U1000}
    \end{subfigure}

    \vspace{0.5em}

    \begin{subfigure}[t]{0.90\textwidth}
        \centering
        \includegraphics[width=\textwidth]
        {Figures/FigureA3a.png}
        \caption{}
        \label{fig:appendix_cp_U1000}
    \end{subfigure}

    \caption{
    Evolution of the wake and associated pressure field at
    $Re=5.00\times10^{4}$ and $Fr=4.08$ for the 10 PPI,
    20 PPI, 60 PPI, and smooth cylinders.
    Columns correspond, from left to right, to 10 PPI, 20 PPI,
    60 PPI, and the smooth cylinder, while the rows correspond to
    approximately $h=1.5$, $1.0$, and $0.5$.
    (a) Nondimensional vorticity, $\omega D/U$, with the in-plane
    velocity vectors superimposed.
    (b) Corresponding reconstructed pressure-coefficient field,
    $C_p$.
    The black contour delineates the clockwise vortex used for the
    quantitative analysis. The white circular region denotes the
    cylinder, and the horizontal dashed line indicates the initially
    undisturbed free-surface level.
    The same colour scale is used for all four surface conditions
    within each panel.
    }
    \label{fig:appendix_U1000}
\end{figure*}

For completeness, the instantaneous vorticity and reconstructed
pressure fields for the remaining Reynolds and Froude numbers are
presented in Figs.~\ref{fig:appendix_U300}--\ref{fig:appendix_U1000}.
The corresponding fields for
$Re=3.50\times10^{4}$ and $Fr=2.00$ are shown in
Fig.~\ref{fig:wake_pressure_evolution_Re35000} in the main text.
As in the main-text figure, the black contour delineates the
clockwise vortex selected for the quantitative analysis.

\section{Evolution of wake circulation at fixed surface condition}

\begin{figure*}[t]
    \centering
    \includegraphics[width=1\textwidth]{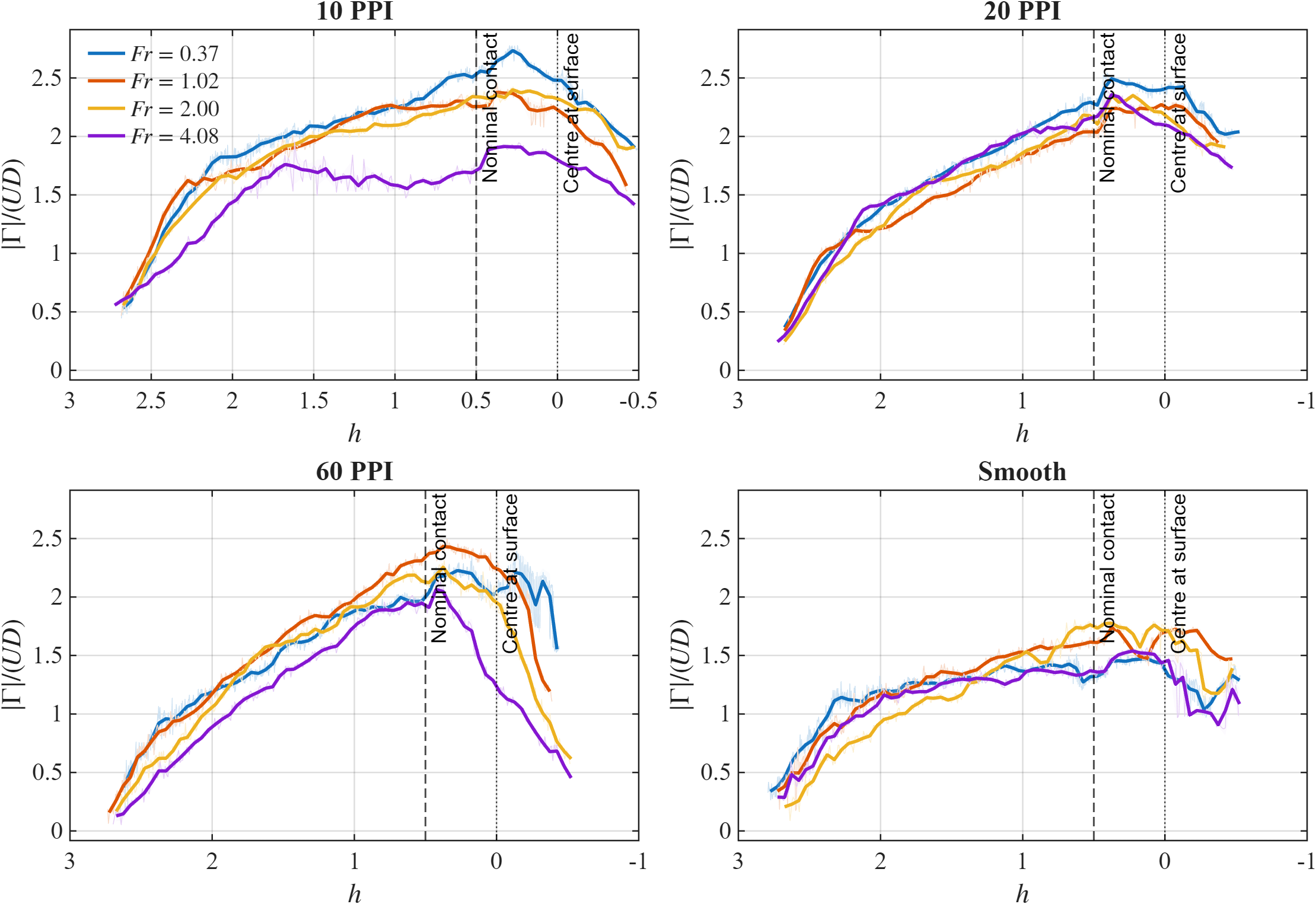}
    \caption{
    Evolution of the nondimensional wake circulation,
    $|\Gamma|/(UD)$, as a function of the cylinder position $h$ for
    \textbf{(a)} 10 PPI,
    \textbf{(b)} 20 PPI,
    \textbf{(c)} 60 PPI, and
    \textbf{(d)} the smooth reference cylinder.
    Each panel compares the four tested operating conditions,
    $Fr=0.37$, 1.02, 2.00, and 4.08.
    The thick curves denote the mean values evaluated in fixed positional bins,
    while the lighter curves show the corresponding instantaneous
    frame-by-frame values.
    The dashed vertical line at $h=0.5$ marks the nominal first-contact
    position, and the dotted vertical line at $h=0$ denotes the position
    at which the cylinder centre reaches the initially undisturbed free surface.
    Since $h$ decreases as the cylinder rises, the evolution is read from
    left to right in the direction of motion.
    }
    \label{fig:supp_gamma_fixed_surface}
\end{figure*}
\nocite{}
\bibliography{aipsamp}% Produces the bibliography via BibTeX.

\end{document}